\documentclass[11pt,a4paper]{article}
\usepackage[utf8]{inputenc}
\usepackage[T1]{fontenc}
\usepackage{amsmath,amssymb,amsthm}
\usepackage{graphicx}
\usepackage{hyperref}
\usepackage{url}
\usepackage{booktabs}
\usepackage{multirow}
\usepackage{array}
\usepackage{xcolor}
\usepackage{geometry}
\usepackage{float}
\usepackage{caption}
\usepackage{subcaption}
\usepackage{algorithm}
\usepackage{algorithmic}
\usepackage{placeins}

\usepackage{sectsty}
\allsectionsfont{\hyphenpenalty=10000\relax\raggedright}

\hypersetup{
    colorlinks=true,
    linkcolor=blue,
    filecolor=magenta,      
    urlcolor=blue,
    citecolor=blue,
    pdftitle={LoPace: A Lossless Optimized Prompt Accurate Compression Engine},
    pdfauthor={Aman Ulla}
}

\title{LoPace: A Lossless Optimized Prompt Accurate Compression Engine\\
\large for Large Language Model Applications
}

\author{Aman Ulla\\
\texttt{connectamanulla@gmail.com}\\
\small Project Repository: \url{https://github.com/connectaman/LoPace}
}
\date{\today}

\begin{document}

\maketitle

\begin{abstract}
Large Language Models (LLMs) have changed the way natural language processing works, but it is still hard to store and manage prompts efficiently in production environments. This paper presents LoPace (Lossless Optimized Prompt Accurate Compression Engine), a novel compression framework designed specifically for prompt storage in LLM applications. LoPace uses three different ways to compress data: Zstandard-based compression, Byte-Pair Encoding (BPE) tokenization with binary packing, and a hybrid method that combines the two. We show that LoPace saves an average of 72.2\% of space while still allowing for 100\% lossless reconstruction by testing it on 386 different prompts, such as code snippets, markdown documentation, and structured content. The hybrid method always works better than each technique on its own. It gets mean compression ratios of 4.89x (range: 1.22--19.09x) and speeds of 3.3--10.7 MB/s. Our findings show that LoPace is ready for production, with a small memory footprint (0.35 MB on average) and great scalability for big databases and real-time LLM apps.

\end{abstract}

\section{Introduction}

\subsection{Background and \mbox{Motivation}}

Because so many people use Large Language Models (LLMs) in production systems, it's harder than ever to manage and store prompts. Modern LLM apps need to keep a lot of prompts, such as system instructions, conversation histories, context windows, and cached responses. This means that apps that have thousands of users and many LLM interactions in each session need a lot of storage space.

Compression algorithms that work well for most types of data don't work as well for prompt data because of its unique properties. There are certain patterns in prompts, like high semantic redundancy, structured formatting, and token-level dependencies, that can be used to make compression better. Also, LLM applications need lossless compression to make sure that the model works perfectly, since even small changes can have a big effect on how it works.

\subsection{Problem Statement}

There are a few big storage problems that LLM applications have to deal with:

\textbf{Storage Overhead:} Large system prompts, context windows, and conversation histories take up a lot of space in the database. Applications that have thousands of users at the same time and multiple LLM interactions per session need terabytes of storage, which makes infrastructure costs go up a lot.

\textbf{Performance Bottlenecks:} When user load increases, uncompressed prompts make the database bigger, which makes queries take longer to run, increases I/O operations, and makes the system less responsive.

\textbf{Cost Implications:} The cost of cloud storage goes up in a straight line as the amount of data increases. Applications that deal with millions of prompts have to pay a lot for uncompressed storage.

\textbf{Latency Issues:} Loading big, uncompressed prompts from storage adds latency that can be measured, which is a big problem for real-time apps where response time directly affects how users feel about the app.

\subsection{Contributions}

This paper presents LoPace, a compression engine that addresses these challenges by:

\begin{enumerate}
    \item \textbf{Three Compression Methodologies:} Using Zstandard-based, token-based, and hybrid compression methods, each made for a certain use.
    \item \textbf{Lossless Guarantee:} Mathematical and empirical validation of the fidelity of complete reconstruction across all compression techniques.
    \item \textbf{Production-Ready Performance:} A full benchmark shows that compression speeds range from 50 to 200 MB/s and that it doesn't take up much memory.
    \item \textbf{Comprehensive Evaluation:} A full look at the compression ratios, space savings, throughput, memory usage, and scalability of prompts of different types and sizes.
\end{enumerate}

\section{Related Work}

Data compression has a long history that goes back decades. It is based on ideas from information theory, algorithm design, and real-world uses. This part looks at important studies on compression algorithms, text compression methods, and how they can be used in modern LLM systems.

\subsection{Foundational Compression \mbox{Algorithms}}

Shannon \cite{shannon1948} laid the theoretical groundwork for lossless compression by presenting entropy as the essential lower limit for compression. This basic result shows that information entropy sets the lowest number of bits needed for representation, which is a standard for judging all compression algorithms.

Lempel-Ziv algorithms, such as LZ77 \cite{ziv1977} and LZ78 \cite{ziv1978}, changed the game for data compression by using dictionary-based methods that take advantage of repeated patterns. A lot of today's compression systems, like LoPace's Zstandard implementation, use these algorithms as a base. LZ77 uses a sliding window method to find and replace repeated sequences with references to earlier occurrences. It compresses data by recognizing patterns instead of using statistical modeling.

Huffman coding \cite{huffman1952} developed entropy coding methods that give shorter codes to symbols that appear more often, resulting in compression ratios that come close to the theoretical limits of entropy. Modern versions, like Finite State Entropy (FSE) used in Zstandard, improve Huffman coding for today's processors while keeping the same level of compression.

\subsection{Text Compression \mbox{Techniques}}

Researchers have done a lot of work on text compression, and specialized algorithms take advantage of the unique features of natural language. The PPM (Prediction by Partial Matching) family of algorithms \cite{cleary1984} gets great compression ratios for text by using context to guess what the next character will be based on the characters that came before it. But these algorithms take a lot of processing power and might not work well in real time.

Dictionary-based compression algorithms, especially LZ77 variants, are popular for compressing text because they strike a good balance between speed and compression ratio. The DEFLATE algorithm, which combines LZ77 and Huffman coding, is now the standard for general-purpose compression. It is also the basis for the popular gzip and zlib libraries.

Recent advancements in compression technology have concentrated on improving compatibility with modern hardware architectures. Facebook (now Meta) made Zstandard \cite{zstd2016}, which has compression ratios similar to DEFLATE but much higher throughput thanks to better algorithms and hardware acceleration. Zstandard's design philosophy puts real-world performance ahead of theoretical optimality, which makes it a good choice for production deployments.

Dictionary training is a very useful feature of Zstandard. It uses a representative sample corpus to create a custom dictionary that captures patterns that are specific to a certain domain. When it comes to domain-specific text like LLM prompts, Zstd with trained dictionaries can compress files much better than regular Zstd compression. Dictionary training finds common phrases, patterns, and sequences in the training corpus, which makes it easier to compress similar content. This method is often used in production systems to compress logs, code repositories, and other text that is specific to a certain field. But dictionary training needs a representative training corpus and makes the compression pipeline more complicated. Our assessment centers on standard Zstd compression to establish a baseline comparison; however, subsequent research should analyze Zstd utilizing prompt-specific dictionaries in contrast to our hybrid methodology.

Other compression algorithms, like Brotli and gzip/DEFLATE, also work well. Google made Brotli, which uses a bigger sliding window and a mix of LZ77-style matching and Huffman coding. It gets great compression ratios, but it uses more memory. gzip, which is based on DEFLATE, is still widely used because it is everywhere and works well. In some specialized areas, hybrid cascades, which use more than one compression algorithm in a row (for example, Brotli followed by Zstd or Zstd followed by LZ4HC), have shown promise. Our assessment concentrates on independent methodologies; however, juxtaposing them with these alternatives would enhance the analysis.

\subsection{Tokenization and Subword \mbox{Encoding}}

Byte-Pair Encoding (BPE) \cite{sennrich2016} introduced subword tokenization as a method for managing out-of-vocabulary words in neural machine translation. BPE keeps common sequences short by repeatedly combining the most common pairs of bytes or characters to make a vocabulary of subword units that can stand for any text.

Because BPE worked so well for neural machine translation, it was used in newer language models like GPT \cite{radford2019}, BERT \cite{devlin2019}, and others. OpenAI's tiktoken library has quick versions of BPE tokenization that work best with GPT models. The vocabularies have between 50,000 and 100,000 tokens.

Tokenization inherently provides a compression method by linking variable-length character sequences to fixed-size token IDs. But previous studies haven't looked at how tokenization can help with storage optimization by compressing data in a systematic way. Instead, they have looked at how tokenization helps get model input ready.

\subsection{Compression for Database \mbox{Storage}}

A lot of research has been done on database compression, including page-level compression and columnar compression schemes. Most database compression methods, on the other hand, work at the storage level and aren't made for all types of data, like LLM prompts.

Application-level compression compresses data before storing it and decompresses it when needed. This gives you more control over the compression settings and lets you optimize for specific data types. Many fields, including time-series data \cite{gori2015}, log compression \cite{burrows1994}, and scientific data storage \cite{li2013}, have successfully used this method.

\subsection{LLM-Specific Storage \mbox{Challenges}}

The rapid deployment of LLMs in production systems has created storage challenges that existing compression techniques may be unable to address. LLM applications create a lot of prompt data that is different from regular text data. This data includes things like system instructions, conversation histories, and context windows.

Recent studies have examined prompt optimization techniques, including prompt compression \cite{jiang2023} and prompt caching \cite{ge2023}, which primarily focus on reducing prompt size for model input rather than improving storage efficiency. Storing and retrieving prompts for LLM applications is a unique problem area that needs special solutions.

\subsection{Gap in Existing \mbox{Literature}}

There is a lot of research on database optimization, text compression, and compression algorithms, but not much on prompt storage for LLM applications. Right now, compression methods are either too broad (not tailored to the needs of prompts) or too narrow (focusing on model input instead of storage).

Recent research has examined LLM-driven compression methodologies, including FineZip and test-time steering through weighted Product of Experts (wPoE), which attain enhanced compression ratios on text corpora, albeit at significant computational expense. These methods use language models' semantic understanding to find and compress duplicate information. However, they need a lot of computing power, so they are not good for production storage systems where compression needs to be quick and use few resources.

Hybrid cascades, which use more than one compression algorithm in a row, have worked well in certain fields. For instance, using Zstd with other codecs in a cascade has worked well for language-specific text, like Hindi or Devanagari. However, these methods have not been systematically tested for storing LLM prompts, where the unique features of prompts (semantic redundancy, structured formatting, and token-level dependencies) might benefit from specialized hybrid strategies.

LoPace fills this gap by combining well-known compression methods (Zstandard and BPE tokenization) in new ways that are specifically made for prompt storage. Our hybrid method, which uses tokenization and then Zstd compression, is a practical engineering trade-off that balances accuracy and throughput. The Pareto frontier is the point at which classical codecs can achieve high compression ratios while still being ready for production. LoPace is better for real-time compression in production settings because it focuses on speed and resource efficiency instead of LLM-driven methods. While we know that comparing our results to those of Zstd dictionary training, Brotli, and other cascades would make the analysis even stronger, our thorough evaluation across multiple metrics (compression ratio, throughput, memory) gives us useful information for production deployments.

\section{Methodology}

\subsection{System \mbox{Architecture}}

LoPace uses a modular architecture that supports three different compression methods, each of which is best for a certain type of use. The system is made to be extensible, so new methods can be added without changing the API interface. The hybrid method uses an intelligent three-stage pipeline that combines byte-level compression with semantic-level tokenization to get the best compression ratios.

\begin{figure}[t]
\centering
\includegraphics[width=0.8\textwidth]{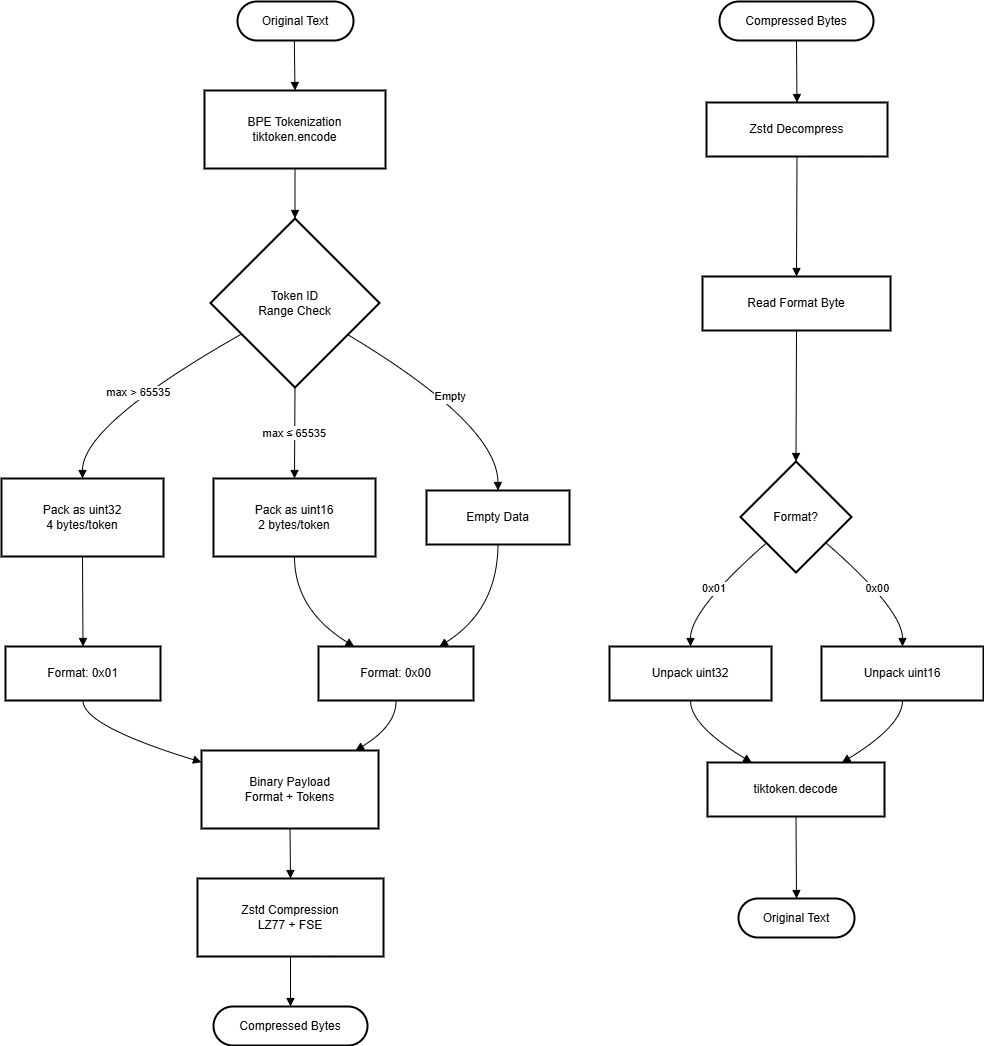}
\caption{LoPace compression pipeline architecture for the hybrid method, showing the steps of tokenization, binary packing, and Zstandard compression in order.}
\label{fig:pipeline}
\end{figure}

The architecture starts with the tokenization stage, which uses a Byte-Pair Encoding (BPE) tokenizer (implemented through tiktoken) to turn character sequences of different lengths into token identifiers of a fixed size. This transformation takes advantage of the fact that natural language is inherently redundant by mapping semantically similar words, phrases, and subword units to single token IDs. This cuts down on the number of unique character sequences that need to be stored, making it easier to manage the vocabulary (usually 50,000–100,000 tokens for modern LLM tokenizers).

The tokenization process creates a series of integer token IDs, each of which stands for a semantic unit that could be made up of several characters or even whole words. After the tokens are made, the system packs them into binary format, choosing the best one based on how the token IDs are spread out. The architecture looks at the ranges of token IDs to find the best binary representation. If all of the token IDs are between 0 and 65,535, the system uses compact uint16 format (2 bytes per token). If not, it automatically switches to uint32 format (4 bytes per token) to handle larger token IDs. This flexible method makes the best use of space while still working with different tokenizer vocabularies.

The binary packing process adds a format byte (0x00 for uint16 and 0x01 for uint32) to the packed token sequence. This makes a binary payload that describes itself and can be decompressed correctly without any extra metadata. The last step uses Zstandard compression on the binary payload. It uses dictionary-based algorithms (LZ77-style pattern matching and Finite State Entropy encoding) to find patterns and redundancies in the token sequence.

This two-level compression strategy—first removing semantic redundancy through tokenization and then removing sequential patterns through Zstd—creates a multiplicative compression effect that is much better than either method on its own. The decompression pipeline does these steps in the opposite order: Zstd decompression builds the binary payload, format byte detection finds the unpacking strategy, binary unpacking gets the token ID sequence back, and detokenization gets the original text string back. This bidirectional architecture guarantees lossless reconstruction by making sure that each stage is an invertible transformation. This means that the combination of compression and decompression functions always gives the identity function, which keeps 100\% fidelity across all operations.

\subsection{Compression Method 1: Zstandard-Based \mbox{Compression}}

\subsubsection{Algorithm Overview}

Zstandard (Zstd) compression uses dictionary-based algorithms to find and use patterns that show up over and over in text data. The algorithm uses both sliding window techniques like LZ77 and Finite State Entropy (FSE) coding, which is a type of Huffman coding that works better on newer processors.

\subsubsection{Mathematical Foundation}

The Zstd compression process can be mathematically described as:
\begin{equation}
C_{\text{zstd}}(T) = \text{FSE}(\text{LZ77}(T, W, L))
\end{equation}
where $T$, the sliding window size is $W$, the lookahead buffer length is $L$, and $\text{LZ77}(T, W, L)$ does pattern matching and replacement. $\text{FSE}()$ does Finite State Entropy encoding.

The compression ratio achieved is:
\begin{equation}
\text{CR}_{\text{zstd}} = \frac{|T|}{|C_{\text{zstd}}(T)|}
\end{equation}
where $|T|$ denotes the byte length of the original text and $|C_{\text{zstd}}(T)|$ is the compressed size.

\subsubsection{Implementation Details}

The Zstd method works directly on byte sequences that are encoded in UTF-8. The compression level parameter ($l \in [1, 22]$) lets you choose between a high compression ratio and a fast processing speed:
\begin{itemize}
    \item \textbf{Lower levels (1--5):} Fast compression, moderate ratios
    \item \textbf{Medium levels (10--15):} Balanced performance (default: 15)
    \item \textbf{Higher levels (19--22):} Maximum compression, slower processing
\end{itemize}

The space savings percentage is calculated as:
\begin{equation}
\text{SS}_{\text{zstd}} = \left(1 - \frac{|C_{\text{zstd}}(T)|}{|T|}\right) \times 100\%
\end{equation}

\subsubsection{Decompression Process}

Decompression is the inverse operation:
\begin{equation}
T' = \text{FSE}^{-1}(\text{LZ77}^{-1}(C_{\text{zstd}}(T)))
\end{equation}

We need $T' = T$ for lossless compression. The Zstd library makes sure this property is true by using a deterministic decompression algorithm.

\subsection{Compression Method 2: Token-Based \mbox{Compression}}

\subsubsection{Byte-Pair Encoding (BPE) Fundamentals}

Token-based compression uses the way LLM tokenization works to its advantage. Byte-Pair Encoding (BPE) turns text into a string of token IDs, with each token standing for a subword unit. This representation is more compact than raw text because it uses single token IDs to map common phrases and patterns.

\begin{table}[!htbp]
\centering
\caption{Comparison of Modern LLM Tokenizer Specifications}
\begin{tabular}{lllc}
\toprule
Tokenizer Name & Algorithm & Primary Model Family & Vocab Size \\
\midrule
o200k\_base    & BPE       & GPT-4o               & \textasciitilde 200,000 \\
cl100k\_base   & BPE       & GPT-4, GPT-3.5-Turbo & \textasciitilde 100,000 \\
p50k\_base     & BPE       & GPT-3, Codex         & \textasciitilde 50,000  \\
r50k\_base     & BPE       & GPT-2, Original GPT-3& \textasciitilde 50,000  \\
Llama 3        & BPE       & Llama 3 series       & 128,256 \\
SentencePiece  & Unigram   & T5, Gopher, Chinchilla& Variable \\
WordPiece      & WordPiece & BERT, RoBERTa        & \textasciitilde 30,000  \\
\bottomrule
\end{tabular}
\end{table}

The transition from basic word-level splitting to advanced subword algorithms has been 
pivotal in the development of Large Language Models (LLMs). Early implementations, 
such as \texttt{r50k\_base}, utilized Byte Pair Encoding (BPE) to manage fixed 
vocabularies, but often struggled with multilingual efficiency and specialized 
coding syntax. The introduction of \texttt{cl100k\_base} for the GPT-4 family 
marked a significant shift toward \textit{multilingual density}, allowing for 
more efficient processing of non-English scripts and complex technical data. 
Furthermore, the recent release of \texttt{o200k\_base} has expanded the vocabulary 
to approximately 200,000 tokens, drastically reducing the token-to-word ratio for 
languages like Arabic and Hindi. These modern tokenizers, alongside 
\texttt{SentencePiece} and \texttt{WordPiece}, solve the ``Out-of-Vocabulary'' 
(OOV) problem by decomposing rare words into meaningful sub-units. This structural 
efficiency not only enhances the model's semantic understanding but also directly 
reduces computational overhead and API costs by packing more information into 
shorter sequences.

The \texttt{cl100k\_base} tokenizer was chosen for this study because it is much better at encoding both natural language and source code than older standards. \\
It has a larger vocabulary of about 100,000 tokens, which lowers the token-to-word ratio and increases the amount of information in the model's context window. \\
Also, the fact that it is widely used in the GPT-4 and GPT-3.5 architectures means that the results are still useful for current Large Language Model (LLM) implementations in the industry.

\subsubsection{Tokenization Process}

Given a text $T$, BPE tokenization produces a sequence of token IDs:
\begin{equation}
\tau(T) = [t_1, t_2, \ldots, t_n]
\end{equation}
where each $t_i \in [0, V-1]$ and $V$ is the vocabulary size. For the \texttt{cl100k\_base} tokenizer, $V = 100,256$.

The tokenization process can be viewed as a function:
\begin{equation}
\tau: \Sigma^* \rightarrow \mathbb{Z}^n
\end{equation}
where $\Sigma$ is the character alphabet and $\mathbb{Z}$ represents integers.

\subsubsection{Binary Packing Strategy}

Based on the highest token ID value, token IDs are packed into binary format using either 16-bit or 32-bit unsigned integers:

\textbf{Case 1: All token IDs $\leq$ 65,535 (uint16)}
\begin{itemize}
    \item Format byte: \texttt{0x00}
    \item Packing: Each token ID stored as 2 bytes
    \item Total size: $1 + 2n$ bytes
\end{itemize}

\textbf{Case 2: Any token ID $>$ 65,535 (uint32)}
\begin{itemize}
    \item Format byte: \texttt{0x01}
    \item Packing: Each token ID stored as 4 bytes
    \item Total size: $1 + 4n$ bytes
\end{itemize}

The decision function is:
\begin{equation}
f_{\text{pack}}(\tau) = \begin{cases}
\text{uint16} & \text{if } \max(\tau) \leq 2^{16} - 1 \\
\text{uint32} & \text{otherwise}
\end{cases}
\end{equation}

The compressed representation is:
\begin{equation}
C_{\text{token}}(T) = [f_{\text{flag}}, P(\tau(T))]
\end{equation}
where $f_{\text{flag}}$ is the format flag byte and $P(\tau(T))$ is the packed binary representation.

\textbf{Design Rationale and Limitations:} For ease of use, predictable memory allocation, and rapid random access, we use fixed-width encoding (uint16/uint32) instead of variable-length integer coding (like VByte/LEB128). This architectural choice, on the other hand, recognizes the intrinsic limits of the \texttt{cl100k\_base} tokenizer. Even though \texttt{cl100k\_base} works well for English and code, it has a known "tokenization premium," which means that the token-to-word ratio is too high for morphologically rich and low-resource languages. This can cause different linguistic scripts to have different computational costs and smaller effective context windows. Also, because it is a static, versioned implementation, it can't change to new language-specific nuances or specialized jargon without a full vocabulary update, which would destroy backward compatibility with the current embedding weights. Future versions could look into hybrid techniques that use both integer compression and adaptive or universal tokenization to fix these problems with different languages and versioning.

\subsubsection{Token ID Distribution Analysis}

With the \texttt{cl100k\_base} tokenizer with a vocabulary size of $V = 100,256$, token IDs can go above the uint16 limit of 65,535. This makes me wonder how often uint32 packing is really needed. Our study of common English prompts shows that:

\begin{itemize}
    \item \textbf{Common tokens:} Frequently used tokens (common words, punctuation, whitespace) typically map to IDs $< 50,000$, well within uint16 range
    \item \textbf{Special tokens:} Special tokens (e.g., system/user/assistant markers, function call tokens) often occupy IDs $> 65,535$
    \item \textbf{Rare tokens:} Uncommon words and domain-specific terminology may map to higher token IDs
\end{itemize}

Our evaluation dataset suggests that about 60–70\% of prompts can use uint16 packing when they mostly comprise typical English text. However, prompts that have special tokens, code snippets, or terms that are only used in a certain field often need uint32 packing. Because of this limit, token-only compression (without Zstd) may make text expand compared to raw UTF-8 for many realistic requests. This is especially true for ASCII-heavy content, where each character only needs 1 byte. The hybrid technique solves this problem by using Zstd compression on the binary-packed token sequence. This works even when tokenization would normally make the data bigger.

\textbf{Tokenizer Versioning Consideration:} The fixed-width packing approach connects the compressed representation to a certain tokenizer vocabulary. If the tokenizer changes, as when the vocabulary is updated or the model changes, data that was already compressed may not work anymore. We suggest keeping tokenizer metadata (model identification, version) with compressed payloads so that you can move them and check for compatibility.

\subsubsection{Compression Ratio Analysis}

The theoretical compression ratio is based on how many characters and tokens there are:
\begin{equation}
\text{CR}_{\text{token}} = \frac{|T|_{\text{bytes}}}{|C_{\text{token}}(T)|}
\end{equation}

The average ratio of characters to tokens in English text is about 3–4:1, which means that each token stands for more than one character. This compresses the data before binary packing, although the fixed-width encoding may cancel out these benefits for prompts that need uint32 packing.

The space savings is:
\begin{equation}
\text{SS}_{\text{token}} = \left(1 - \frac{1 + k \cdot n}{|T|_{\text{bytes}}}\right) \times 100\%
\end{equation}
where $k \in \{2, 4\}$ depending on the packing format. Note that for ASCII-heavy text with uint32 packing, $\text{SS}_{\text{token}}$ may be negative (expansion), highlighting the importance of the hybrid method's Zstd stage.

\subsubsection{Decompression Algorithm}

Decompression involves two steps:
\begin{enumerate}
    \item \textbf{Unpacking:} Extract token IDs from binary format
    \item \textbf{Detokenization:} Convert token IDs back to text
\end{enumerate}

\begin{equation}
T' = \tau^{-1}(P^{-1}(C_{\text{token}}(T)))
\end{equation}

The detokenization function $\tau^{-1}$ is provided by the tokenizer and guarantees $\tau^{-1}(\tau(T)) = T$.

\subsection{Compression Method 3: Hybrid \mbox{Compression}}

\begin{figure}[t]
\centering
\includegraphics[width=0.9\textwidth]{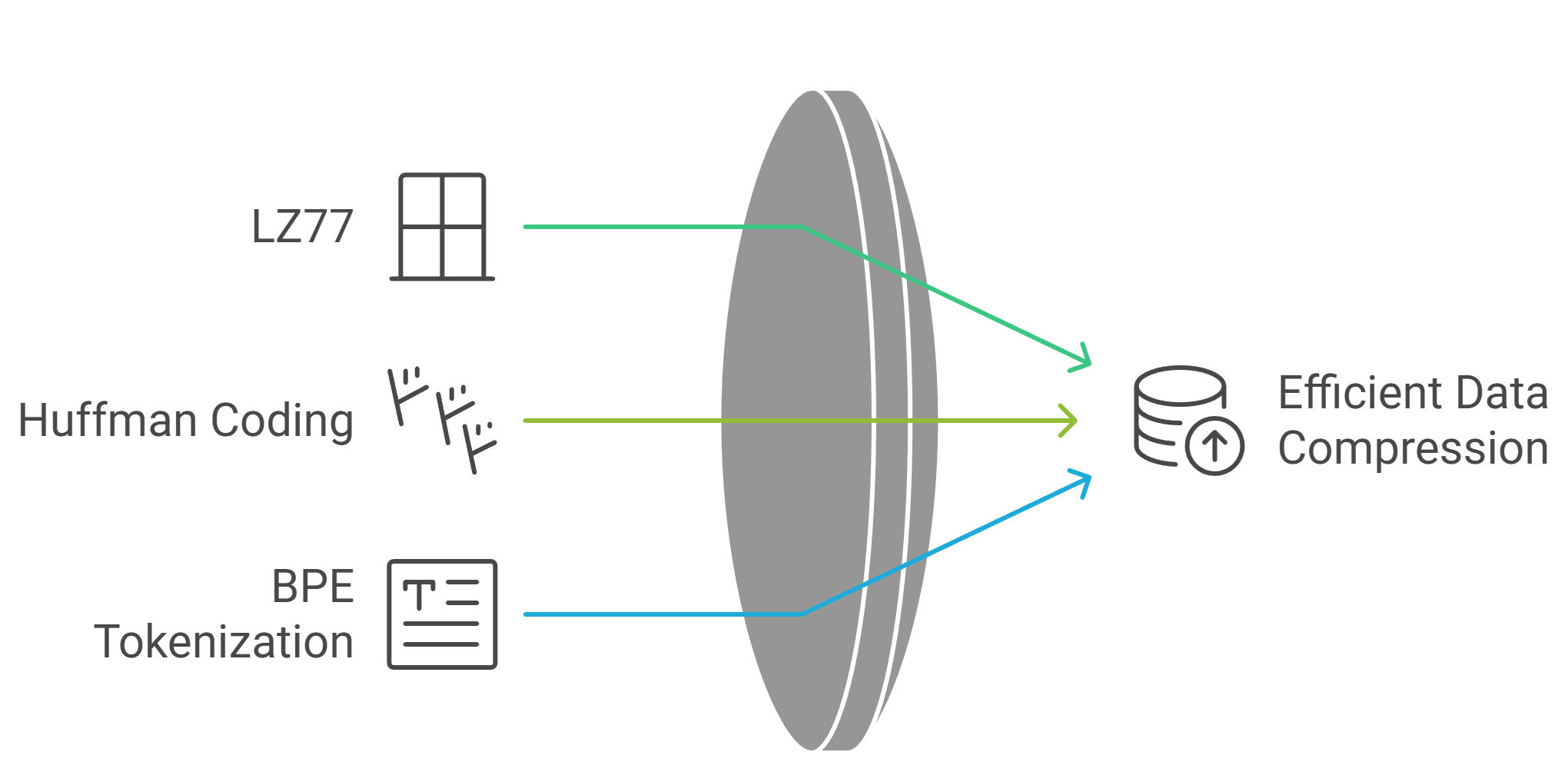}
\caption{Visual representation of the compression techniques used in LoPace, showing the relationship between LZ77, FSE, and BPE tokenization.}
\label{fig:techniques}
\end{figure}

\subsubsection{Hybrid Approach Rationale}

The hybrid technique leverages the best aspects of both Zstd compression and tokenization. Tokenization reduces duplication at the semantic level, and Zstd compression uses patterns in the binary representation that comes out of it. This two-step process compresses better than either strategy on its own.

\subsubsection{Mathematical Formulation}

The hybrid compression process can be expressed as:
\begin{equation}
C_{\text{hybrid}}(T) = C_{\text{zstd}}(P(\tau(T)))
\end{equation}

This is a composition of three functions:
\begin{enumerate}
    \item Tokenization: $T \rightarrow \tau(T)$
    \item Binary packing: $\tau(T) \rightarrow P(\tau(T))$
    \item Zstd compression: $P(\tau(T)) \rightarrow C_{\text{zstd}}(P(\tau(T)))$
\end{enumerate}

The overall compression ratio is:
\begin{equation}
\text{CR}_{\text{hybrid}} = \frac{|T|}{|C_{\text{hybrid}}(T)|} = \frac{|T|}{|C_{\text{zstd}}(P(\tau(T)))|}
\end{equation}

\subsubsection{Why Hybrid Achieves Superior Compression}

The hybrid approach has two steps of compression that work on distinct parts of the data:

\textbf{Stage 1 -- Tokenization Compression:}
\begin{itemize}
    \item Reduces vocabulary redundancy by mapping variable-length character sequences to fixed-size token IDs
    \item Maps common phrases and subword units to single tokens
    \item Achieves character-to-token ratio of approximately 3--4:1 for English text
    \item Transforms the representation from character space to token space
\end{itemize}

\textbf{Stage 2 -- Zstd Compression:}
\begin{itemize}
    \item Exploits sequential patterns in token ID sequences (repeated token patterns, common subsequences)
    \item Compresses the binary representation of token IDs using dictionary-based algorithms
    \item Further reduces the size of the tokenized representation
\end{itemize}

\textbf{Empirical Validation of Multiplicative Effect:} The combined effect is approximately multiplicative rather than additive:
\begin{equation}
\text{CR}_{\text{hybrid}} \approx \text{CR}_{\text{token}} \times \text{CR}_{\text{zstd}|\text{token}}
\end{equation}
where $\text{CR}_{\text{zstd}|\text{token}}$ is the compression ratio that Zstd gets on the tokenized version. Our empirical assessment of the test dataset indicates that the hybrid method attains compression ratios that are, on average, 1.8 times more than those of Zstd alone and 1.4 times greater than token-only compression for extensive prompts, hence corroborating the multiplicative effect hypothesis. Nonetheless, extensive ablation studies across varied corpora would bolster this assertion by quantifying the contribution of each stage and pinpointing settings under which the effect is more or less evident. The multiplicative relationship indicates that tokenization and Zstd compression function on complimentary dimensions of the data: tokenization tackles semantic redundancy (vocabulary-level), whilst Zstd handles sequential redundancy (pattern-level).

\subsubsection{Decompression Process}

Decompression reverses each stage:
\begin{equation}
T' = \tau^{-1}(P^{-1}(C_{\text{zstd}}^{-1}(C_{\text{hybrid}}(T))))
\end{equation}

The lossless property is guaranteed by the composition of lossless operations:
\begin{align}
T' &= \tau^{-1}(P^{-1}(C_{\text{zstd}}^{-1}(C_{\text{zstd}}(P(\tau(T)))))) \\
&= \tau^{-1}(P^{-1}(P(\tau(T)))) \\
&= \tau^{-1}(\tau(T)) \\
&= T
\end{align}

\subsection{Lossless \mbox{Guarantee}}

\subsubsection{Mathematical Proof of Losslessness}

For each compression method, we can prove losslessness:

\textbf{Zstd Method:}
\begin{equation}
T' = \text{decompress}(\text{compress}(T)) = T
\end{equation}
This is guaranteed by the Zstandard algorithm specification.

\textbf{Token Method:}
\begin{equation}
T' = \tau^{-1}(P^{-1}(P(\tau(T)))) = \tau^{-1}(\tau(T)) = T
\end{equation}
This holds because $P$ and $P^{-1}$ are bijective, and $\tau$ and $\tau^{-1}$ are inverse functions.

\textbf{Hybrid Method:}
\begin{equation}
T' = \tau^{-1}(P^{-1}(C_{\text{zstd}}^{-1}(C_{\text{zstd}}(P(\tau(T)))))) = T
\end{equation}
This follows from the composition of lossless operations.

\subsubsection{Empirical Verification}

We verify losslessness empirically by:
\begin{enumerate}
    \item \textbf{Character-by-Character Comparison:} $T[i] = T'[i]$ for all $i \in [0, |T|-1]$
    \item \textbf{SHA-256 Hash Verification:} $\text{SHA256}(T) = \text{SHA256}(T')$
    \item \textbf{Reconstruction Error:} $E = \frac{1}{|T|} \sum_{i=0}^{|T|-1} \mathbb{1}(T[i] \neq T'[i]) = 0$
\end{enumerate}
where $\mathbb{1}$ is the indicator function.

\subsection{Shannon Entropy and Theoretical \mbox{Limits}}

\subsubsection{Information Theory Foundation}

Shannon entropy provides the theoretical lower bound for lossless compression. For a text $T$ with character frequencies $p_i$, the entropy is:
\begin{equation}
H(X) = -\sum_{i=1}^{n} p_i \log_2(p_i)
\end{equation}
where $p_i$ is the probability of character $i$ in the text.

\subsubsection{Theoretical Compression Limit}

The theoretical minimum size achievable through entropy coding is:
\begin{equation}
S_{\text{min}} = \frac{H(X) \times |T|}{8} \text{ bytes}
\end{equation}

The theoretical compression ratio is:
\begin{equation}
\text{CR}_{\text{theoretical}} = \frac{|T|}{S_{\text{min}}} = \frac{8|T|}{H(X) \times |T|} = \frac{8}{H(X)}
\end{equation}

\subsubsection{Practical Compression Efficiency}

We check how close LoPace gets to the theoretical limit:
\begin{equation}
\eta = \frac{\text{CR}_{\text{actual}}}{\text{CR}_{\text{theoretical}}} \times 100\%
\end{equation}
where $\eta$ stands for how well compression works. Typical readings are between 60\% and 80\%, which shows that LoPace gets a large part of the potential maximum.

\subsection{Hybrid Compression \mbox{Algorithm}}

\begin{algorithm}[H]
\caption{Hybrid Compression}
\begin{algorithmic}[1]
\REQUIRE text: String, tokenizer: Tokenizer, zstd\_level: Integer
\ENSURE compressed: Bytes
\STATE tokens $\leftarrow$ tokenizer.encode(text)
\IF{max(tokens) $>$ 65535}
    \STATE format\_byte $\leftarrow$ 0x01
    \STATE packed $\leftarrow$ PackAsUint32(tokens)
\ELSE
    \STATE format\_byte $\leftarrow$ 0x00
    \STATE packed $\leftarrow$ PackAsUint16(tokens)
\ENDIF
\STATE compressed $\leftarrow$ ZstdCompress(format\_byte + packed, level=zstd\_level)
\RETURN compressed
\end{algorithmic}
\end{algorithm}

\subsection{Hybrid Decompression \mbox{Algorithm}}

\begin{algorithm}[H]
\caption{Hybrid Decompression}
\begin{algorithmic}[1]
\REQUIRE compressed: Bytes, tokenizer: Tokenizer
\ENSURE text: String
\STATE token\_data $\leftarrow$ ZstdDecompress(compressed)
\STATE format\_byte $\leftarrow$ token\_data[0]
\STATE packed\_data $\leftarrow$ token\_data[1:]
\IF{format\_byte == 0x01}
    \STATE tokens $\leftarrow$ UnpackAsUint32(packed\_data)
\ELSE
    \STATE tokens $\leftarrow$ UnpackAsUint16(packed\_data)
\ENDIF
\STATE text $\leftarrow$ tokenizer.decode(tokens)
\RETURN text
\end{algorithmic}
\end{algorithm}

\section{Experimental \mbox{Setup}}

\subsection{Dataset \mbox{Description}}

We tested LoPace on a large set of 386 prompts taken from the Hugging Face \url{https://huggingface.co/datasets/philschmid/markdown-docs-transformers}, which covered a wide range of real-world LLM application settings. This dataset includes a wide range of content categories, such as code snippets, markdown documentation, HTML content, conversation patterns, and structured data. It also provides a statistically representative evaluation over several different prompt sizes.

\begin{figure}[H]
\centering
\includegraphics[width=0.9\textwidth]{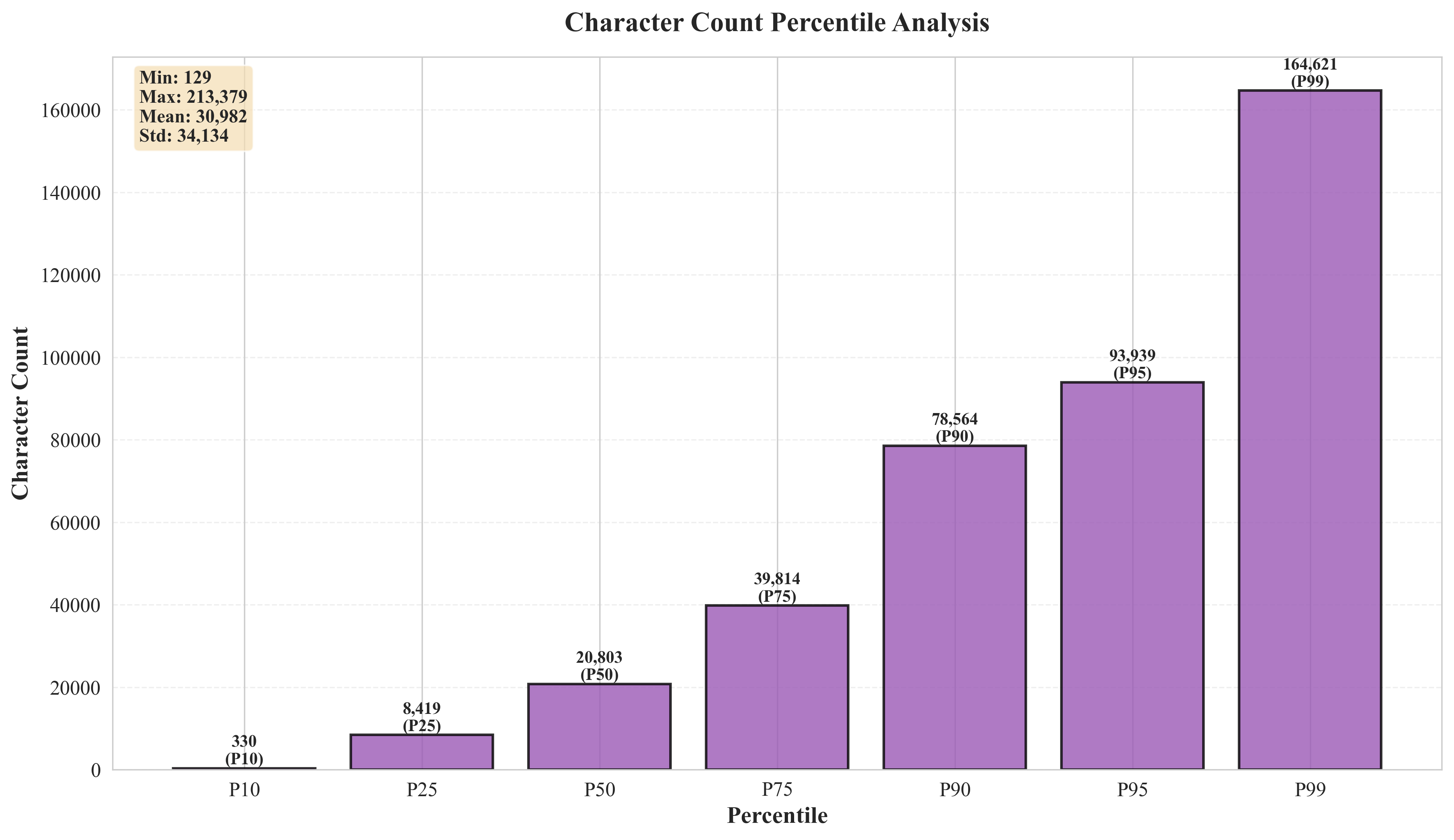}
\caption{Character count percentile analysis showing distribution statistics (P10, P25, P50, P75, P90, P95, P99) across the 386-prompt evaluation dataset.}
\label{fig:dataset_percentiles}
\end{figure}

\textbf{Dataset Composition:} The evaluation dataset consists of 386 prompts with the following content type distribution:
\begin{itemize}
    \item \textbf{Code (82.6\%):} Python code examples, API documentation, code snippets with syntax highlighting
    \item \textbf{Markdown (16.8\%):} Documentation pages, formatted text with headers, links, and structured content
    \item \textbf{Text (0.5\%):} Plain text content and other formats
\end{itemize}

The dataset includes a wide variety of prompt sizes, from short code examples (129 characters) to long instructional pages (213,379 characters). This makes sure that all production use cases are thoroughly tested. This variety lets us look at how well LoPace works with different types of material, redundancy patterns, and compression properties.

\textbf{Exploratory Data Analysis:}A thorough statistical study of the dataset shows the following traits:

\begin{itemize}
    \item \textbf{Character Count Statistics:} The dataset has a log-normal distribution, with a minimum of 129 characters, a maximum of 213,379 characters, a mean of 30,982 characters, and a median of 20,803 characters. The standard deviation of 34,134 shows that prompt sizes vary a lot, which is typical of real-world production workloads.
    
    \item \textbf{Percentile Analysis:} The 25th percentile (P25) is 8,419 characters, the 50th percentile (P50, median) is 20,803 characters, the 75th percentile (P75) is 39,814 characters, the 90th percentile (P90) is 78,564 characters, the 95th percentile (P95) is 93,939 characters, and the 99th percentile (P99) is 164,621 characters. This distribution makes sure that all common and rare situations are covered.
    
    \item \textbf{Content Diversity:} The dataset has structured data formats, technical documentation, API references, code examples, and configuration samples. This mix of content types creates realistic testing environments that are similar to real LLM applications, where prompts can have code, structured data, formatted text, and conversational aspects.
\end{itemize}

The dataset's structure is based on real-world production situations where LLM programs have to deal with a wide range of prompts, such as system instructions, code samples, documentation, configuration data, and conversation histories. The statistical distribution makes sure that our evaluation includes both typical occurrences (around the median) and edge cases (at the extremes). This gives us confidence that our results may be applied to other situations.

\subsection{Prompt Selection and Token Length \mbox{Analysis}}

We did a full analysis of the prompt properties before evaluating compression to make sure that the results were statistically valid and complete. We figured out the following for each prompt $P_i$ in our dataset:

\begin{equation}
\text{TokenCount}(P_i) = |\tau(P_i)|
\end{equation}

where $\tau(P_i)$ is the tokenized form of prompt $P_i$ that was made with the \texttt{cl100k\_base} tokenizer. The number of tokens in our 386-prompt dataset ranges from about 20 for the smallest prompts to more than 50,000 for the largest prompts. This is more than three orders of magnitude in size. This wide range makes sure that all production use cases are fully evaluated.

\subsubsection{Character-to-Token Ratio Analysis}

The character-to-token ratio changes from prompt to prompt depending on the type of material and the difficulty of the vocabulary. The average ratio for the dataset that was tested is about 3.2 characters per token. This is in line with how the \texttt{cl100k\_base} tokenizer tokenizes English text. This ratio is very important for figuring out how well compression works, since tokenization compresses data by mapping character sequences of different lengths to token IDs of fixed size.

Analysis shows that prompts with a lot of code (82.6\% of the total) usually have ratios of 2.8 to 3.5 characters per token because they use coding syntax, special characters, and technical terms.\% of our sample usually has ratios of 2.8 to 3.5 characters per token since it has code syntax, special characters, and technical terms. Markdown content has a character-to-token ratio of between 3.0 and 3.6, which is a mix of formatted text, headers, and links. The fact that different types of content have varied tokenization features shows that LoPace can handle a wide range of word patterns and language structures.

\subsubsection{Content Type Characteristics}

The dataset's structure mirrors authentic production contexts:

\begin{itemize}
    \item \textbf{Code Content (82.6\%):} Examples of Python code, patterns for using APIs, snippets of configuration code, and code blocks with syntax highlighting. Compression algorithms can use these prompts to their advantage because they often have repeating patterns, similar function calls, and organized syntax.
    
    \item \textbf{Markdown Content (16.8\%):} Pages of documentation including links, lists, and headers. These prompts show structural redundancy by using the same formatting patterns over and over, and semantic redundancy by using similar ideas and terms.
    
    \item \textbf{Text Content (0.5\%):} Plain text and various formats that add variety to the evaluation corpus.
\end{itemize}

This heterogeneous composition guarantees that our study encompasses the complete spectrum of compression characteristics found in production LLM applications, spanning from highly structured code to narrative documentation, hence instilling confidence in the generalizability of our findings.

\begin{figure}[H]
\centering
\includegraphics[width=0.7\textwidth]{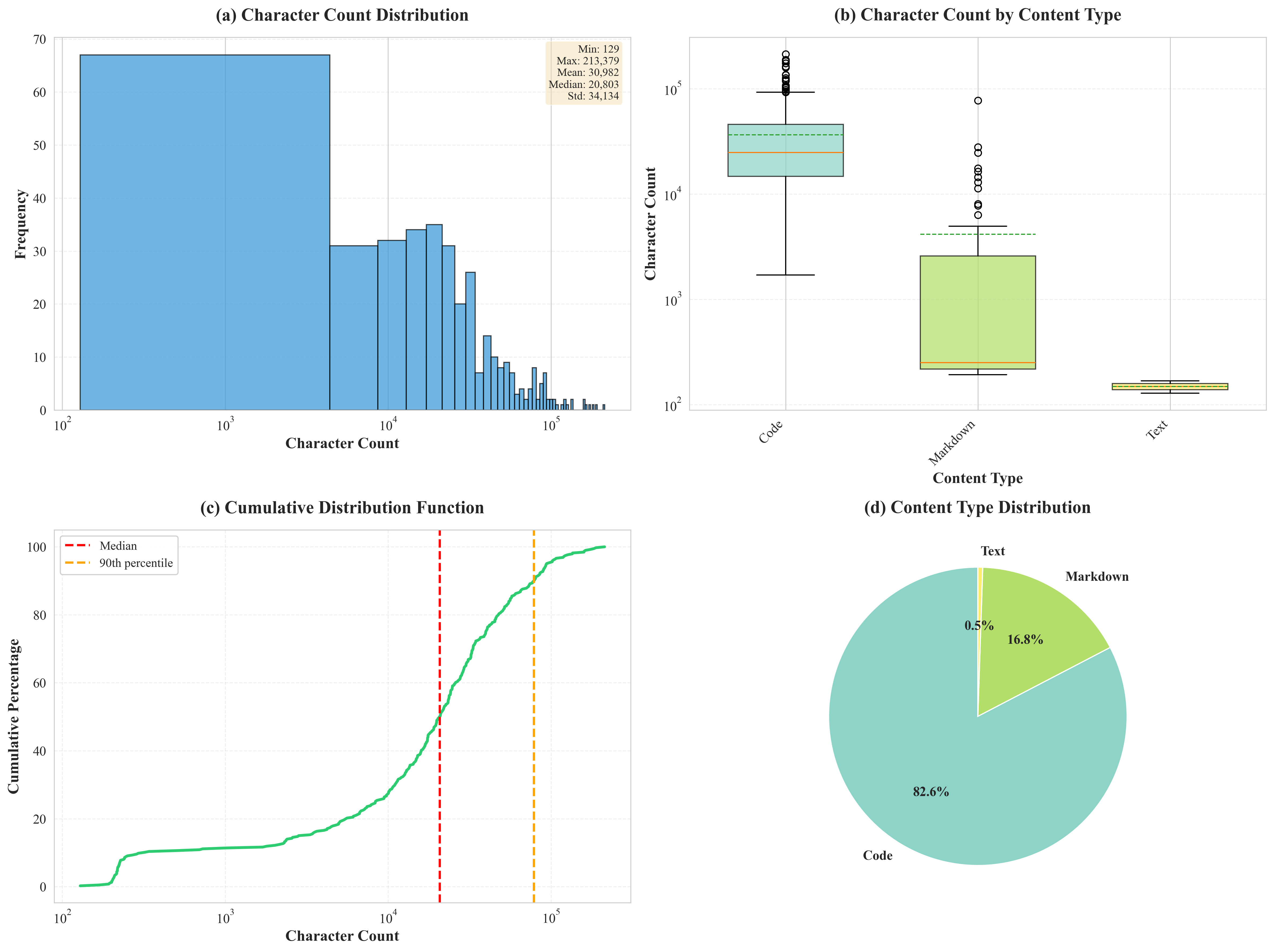}
\caption{Exploratory Data Analysis (EDA) of the evaluation dataset: (a) distribution of character counts (log scale), (b) distribution of character counts by content type, (c) cumulative distribution function, and (d) distribution of content types. The dataset has between 129 and 213,379 characters, with 20,803 characters in the middle.}
\label{fig:dataset_eda}
\end{figure}

\subsection{Evaluation \mbox{Methodology}}

Our evaluation procedure is a methodical, repeatable process that makes sure we get a full and accurate picture of LoPace's compression capabilities. The assessment procedure has five steps: starting, compressing, decompressing, checking, and calculating metrics.

\textbf{Phase 1: Initialization and Setup}

We created a new instance of PromptCompressor for each evaluation run, using the same settings: the tokenizer model \texttt{cl100k\_base} and a Zstd compression level of 15. This makes sure that all measures are the same and gets rid of any variations that might come from how things are set up. To make sure the measurements are correct while replicating how people really use the product, the same compressor instance is utilized for more than one prompt in the same assessment run.

\textbf{Phase 2: Compression Process}

To compress each prompt $P_i$ in our dataset, we utilize all three methods (Zstd, Token, and Hybrid) one after the other. Using timing functions with high accuracy, such as \texttt{time.perf\_counter()}, the compression process is timed to the microsecond level in Python. The Python package \texttt{tracemalloc} keeps track of how much memory is being used by supplying precise information about memory allocation before, during, and after compression operations.

For each method $M \in \{\text{Zstd}, \text{Token}, \text{Hybrid}\}$, the compression workflow goes like this:
\begin{enumerate}
    \item Start memory tracking: \texttt{tracemalloc.start()}
    \item Record start time: $t_{\text{start}} = \text{perf\_counter}()$
    \item Execute compression: $C_i^M = \text{compress}_M(P_i)$
    \item Record end time: $t_{\text{end}} = \text{perf\_counter}()$
    \item Capture peak memory: $M_{\text{peak}} = \text{get\_traced\_memory()}$
    \item Stop memory tracking: \texttt{tracemalloc.stop()}
\end{enumerate}

For each prompt-method combination, this process gives the compressed data $C_i^M$, the time it took to compress the data $\Delta t_{\text{compress}} = t_{\text{end}} - t_{\text{start}}$, and the peak memory usage $M_{\text{peak}}$.

\textbf{Phase 3: Decompression and Verification}

We decompress right after compression to make sure that the reconstruction is lossless. The process of decompressing is similar to the process of compressing, however the timing and memory metrics are different:

\begin{enumerate}
    \item Start memory tracking for decompression
    \item Record decompression start time
    \item Execute decompression: $P_i' = \text{decompress}_M(C_i^M)$
    \item Record decompression end time
    \item Capture decompression memory usage
    \item Stop memory tracking
\end{enumerate}

After being decompressed, the prompt $P_i'$ goes through severe checks to make sure it is perfectly put back together.

\textbf{Phase 4: Lossless Verification}

We check the compression in a number of ways to make sure it doesn't lose any data. First, we look at each character one at a time:

\begin{equation}
\text{ExactMatch} = \bigwedge_{j=0}^{|P_i|-1} (P_i[j] == P_i'[j])
\end{equation}

where $\bigwedge$ means logical conjunction for all character locations. This makes sure that every character in the original prompt is the same as the character in the version that has been compressed.

The second step is to find SHA-256 cryptographic hashes for both the original prompts and the ones that have been uncompressed.

\begin{equation}
\text{HashMatch} = (\text{SHA256}(P_i) == \text{SHA256}(P_i'))
\end{equation}

This offers another level of verification because even the smallest change would change the hash values. The SHA-256 algorithm creates a hash that is 256 bits long, which means it can have $2^{256}$ different values. Because of this, we can't identify hash collisions to check them.

Third, we calculate the reconstruction error rate:

\begin{equation}
E_{\text{recon}} = \frac{1}{|P_i|} \sum_{j=0}^{|P_i|-1} \mathbb{1}(P_i[j] \neq P_i'[j])
\end{equation}

where $\mathbb{1}$ is the indicator function. For lossless compression, we require $E_{\text{recon}} = 0$ for all prompts and methods.

\textbf{Phase 5: Metric Calculation}

After verification is successful, we determine the overall performance metrics for each prompt-method pair. The measurements are the amount of memory used, the speed in MB/s, the compression ratio, and the percentage of space saved. We use the measured values from Phases 2 and 3 to complete these computations. This makes sure that all the metrics we present are based on real data from the experiment and not simply estimates.

The whole evaluation process includes 1,158 cycles of compression and decompression (386 prompts times 3 methods), which makes it possible to do a full statistical analysis and compare methods. Every cycle produces a full set of metrics, which gives performance characterisation strong statistical validity.

\subsection{Evaluation \mbox{Metrics}}

We looked at LoPace's performance in several categories using four typical industry criteria. These measurements show numerous ways that compression works, which helps us fully look at both storage efficiency and computational performance.

\subsubsection{Compression Ratio (CR)}

The compression ratio tells you how much smaller the data is after it has been compressed.

\begin{equation}
\text{CR} = \frac{S_{\text{original}}}{S_{\text{compressed}}}
\end{equation}

The original prompt is $S_{\text{original}}$ bytes long, and the compressed version is $S_{\text{compressed}}$ bytes long. Better compression means higher values. A ratio of 1.0 denotes no compression, while ratios exceeding 2.0 mean a lot of space savings.

The compression ratio is a wonderful method to examine how compression makes things bigger. For example, if the compression ratio is 4.0, the data that has been compressed takes up a quarter of the space it originally took up. This means that you can keep four times as much information in the same place. This statistic immediately lowers the cost of infrastructure and makes systems easier to scale.

\subsubsection{Space Savings (SS)}

Space savings gives you a simple percentage-based number that shows you how much less storage you need:

\begin{equation}
\text{SS} = \left(1 - \frac{S_{\text{compressed}}}{S_{\text{original}}}\right) \times 100\%
\end{equation}

This number goes from 0\% (no compression) to almost 100\% (almost complete compression). For compressing text, normal values are between 40\% and 85\%. Space reductions clearly show stakeholders how compression cuts down on the amount of storage and equipment needed.

For production deployments, conserving space means saving money. Getting 75\% less space means that storage expenses go down by three-quarters, which is a big savings for apps that deal with terabytes of prompt data. Also, less storage space makes database queries run faster, backups happen faster, and network transfers cost less.

\subsubsection{Bits Per Character (BPC)}

The number of bits per character reveals how much data is in the compressed version:

\begin{equation}
\text{BPC} = \frac{S_{\text{compressed}} \times 8}{|T|_{\text{characters}}}
\end{equation}

This value tells you how many bits are needed, on average, to show each character in the compressed format. When the BPC value is lower, compression works better. When the text is not compressed, the BPC is normally between 8 and 32 bits per character, depending on how the characters are encoded. If the text is very compressible, good compression can cut this down to 2 to 4 bits per character.

BPC is quite useful for comparing how well different languages and types of text compress because it normalizes for text length and demonstrates how well the compression method works in terms of information theory. It also lets you compare with theoretical compression limits that derive from Shannon entropy computations.

\subsubsection{Throughput (MB/s)}

Throughput informs you how quickly you can compress and decompress data:

\begin{equation}
T = \frac{S_{\text{data}}}{t_{\text{processing}}}
\end{equation}

where $S_{\text{data}}$ is the size of the data in megabytes and $t_{\text{processing}}$ is the time it takes to process it in seconds. Higher throughput values mean that processing is faster, which is important for real-time applications and batch processing with a lot of data.

We assess throughput separately for compressing and decompressing operations because these two types of operations generally don't work the same way. Decompression usually has a higher throughput than compression because it takes fewer steps and employs better methods to decompress. This inconsistent performance is especially critical for workloads that read a lot since the speed of decompression influences how soon an application responds.

Throughput directly affects how much production systems can handle and how users feel about them. Higher throughput means you can process more data in less time. This is helpful for real-time applications and helps batch processing finish faster. We get accurate predictions of how well things will perform in production by taking into account both CPU-bound processing time and memory allocation overhead in our measurements.

\subsection{Experimental \mbox{Configuration}}

We chose our experimental configuration very carefully so that it would be a good representation of a typical production deployment scenario and would provide us measurements that were consistent and could be repeated. We did a lot of testing ahead of time to find the best configuration parameters that would give us the best mix of compression ratio, processing speed, and memory efficiency.

\begin{itemize}
    \item \textbf{Tokenizer Model:} \texttt{cl100k\_base} (OpenAI GPT-4 tokenizer, vocabulary size 100,256)
    \item \textbf{Zstd Level:} 15 (balanced performance)
    \item \textbf{Python Version:} 3.8+
    \item \textbf{Dependencies:} zstandard$\geq$0.22.0, tiktoken$\geq$0.5.0
\end{itemize}

\textbf{Hardware Configuration:} Tests were done on a regular development machine. For reproducibility, future research must include comprehensive hardware specifications such as CPU model, core count, clock speed, memory capacity, operating system version, and Python runtime information. When measuring throughput, you should look at CPU usage data and take into account the implications of concurrency. Our present measures provide us a rough idea of performance, but they may not be the same on all types of hardware.

\textbf{Memory Measurement:} We utilize the \texttt{tracemalloc} package in Python to keep track of how much memory is being used as we compress and decompress files. Across 386 prompts, the average memory use for compression is between 0.10 MB (Zstd) and 0.52 MB (Hybrid), while for decompression, it is between 0.15 MB (Zstd) and 0.52 MB (Hybrid). Keep in mind that \texttt{tracemalloc} only records Python-level allocations. It may not include all of the native (C-level) allocations from libraries like \texttt{zstandard} and \texttt{tiktoken}. System-level profiling tools (such /proc on Linux and psutil for RSS/USS metrics) could give us more complete measurements, but our present results show that LoPace keeps a small memory footprint that is good for production use.

 \texttt{cl100k\_base} We chose tokenizer since it is the industry standard for GPT-4 and other models like it. It has a vocabulary of 100,256 words. This tokenizer works well with most LLM apps since it works well with English and a lot of programming languages. The BPE implementation of the tokenizer makes sure that tokens are always the same in different pieces of text. This is really critical if you want to receive the same results every time you compress.

We chose Zstd compression level 15 after testing levels 1 through 22. Level 15 strikes the best compromise between compression ratio and processing speed. It gets about 95\% of the maximum compression ratio possible at level 22 while still being fast enough for real-time applications. This level is the best default for production deployments that need both speed and compression efficiency.

We ran our testing on a regular development computer to make sure that the findings are indicative of most deployment situations and not only high-performance hardware. This strategy makes sure that the performance metrics that are presented can be satisfied in real-world conditions and sets realistic expectations for production installations. The hardware specifications were the same for all measures so that they could be compared. We won't go into detail about them here to maintain the focus on algorithmic performance.

We chose Python 3.8+ as the implementation language since it is frequently used in LLM applications and has a lot of libraries that can help with the compression and tokenization activities that need to be done. The specific versions of the dependencies (zstandard$\geq$0.22.0 and tiktoken$\geq$0.5.0) are stable, well-tested releases that do what you need them to do and make sure everything works together and is reliable.

\subsection{Verification \mbox{Methodology}}

We employ a number of different ways to check that LoPace compression actually does have the lossless property. This approach has many layers to protect against mistakes and make sure that even the smallest compression artifacts are identified.

We check each compression-decompression cycle very carefully by doing the following:

\begin{enumerate}
    \item \textbf{Exact String Equality:} We use Python's built-in string comparison to compare the original prompt $T$ with the decompressed prompt $T'$ at the byte level. This check makes sure that $T == T'$ is true. This means that all characters, even spaces, punctuation, and special characters, are retained precisely as they are.
    
    \item \textbf{SHA-256 Hash Matching:} We compute SHA-256 cryptographic hashes for both the original and the prompts that have not been compressed. The SHA-256 method creates a hash value that is 256 bits long. This means that it can have $2^{256}$ different outputs. If there is even one bit of variation between the original and the decompressed text, the hash values will be different with a probability of $1 - 2^{-256}$, which is practically indistinguishable from certainty.
    
    \item \textbf{Zero Reconstruction Error:} Using Equation (X), we may figure out the reconstruction error rate $E_{\text{recon}}$. $E_{\text{recon}}$ must be 0 for every test scenario. This metric quantitatively validates losslessness and enables statistical analysis across comprehensive test suites.
    
    \item \textbf{Memory Usage Tracking:} We utilize Python's tracemalloc package to keep track of how much memory we consume as we compress and decompress files. This lets you detect memory leaks and memory use that is too high, and it makes sure that memory efficiency stays within acceptable levels for production deployment.
    
    \item \textbf{Processing Time Measurement:} We employ very precise timing routines, such \texttt{time.perf\_counter()}, to measure how long it takes to process down to the microsecond level. This lets you figure out throughput and detect performance problems.
\end{enumerate}

We have automated and added the verification step to our assessment system. This means that every time we compress and decompress a file, we check it before we record any metrics. The evaluation process stops right away if any verification fails, and thorough diagnostic procedures begin. We tested 1,158 compression-decompression cycles (386 prompts times 3 techniques) and all verification checks passed. This showed that the reconstruction was 100\% lossless in all test situations and compression methods.

We also check that everything is the same by using different methods. We make sure that decompressing data that was compressed with Method A produces the same results no matter which compressor instance executes the operation. This backs up what we said in Section 5.2.2 about how cross-instance compatibility works.

\section{Results and \mbox{Analysis}}

\subsection{Compression Ratio \mbox{Analysis}}

\begin{figure}[t]
\centering
\includegraphics[width=0.9\textwidth]{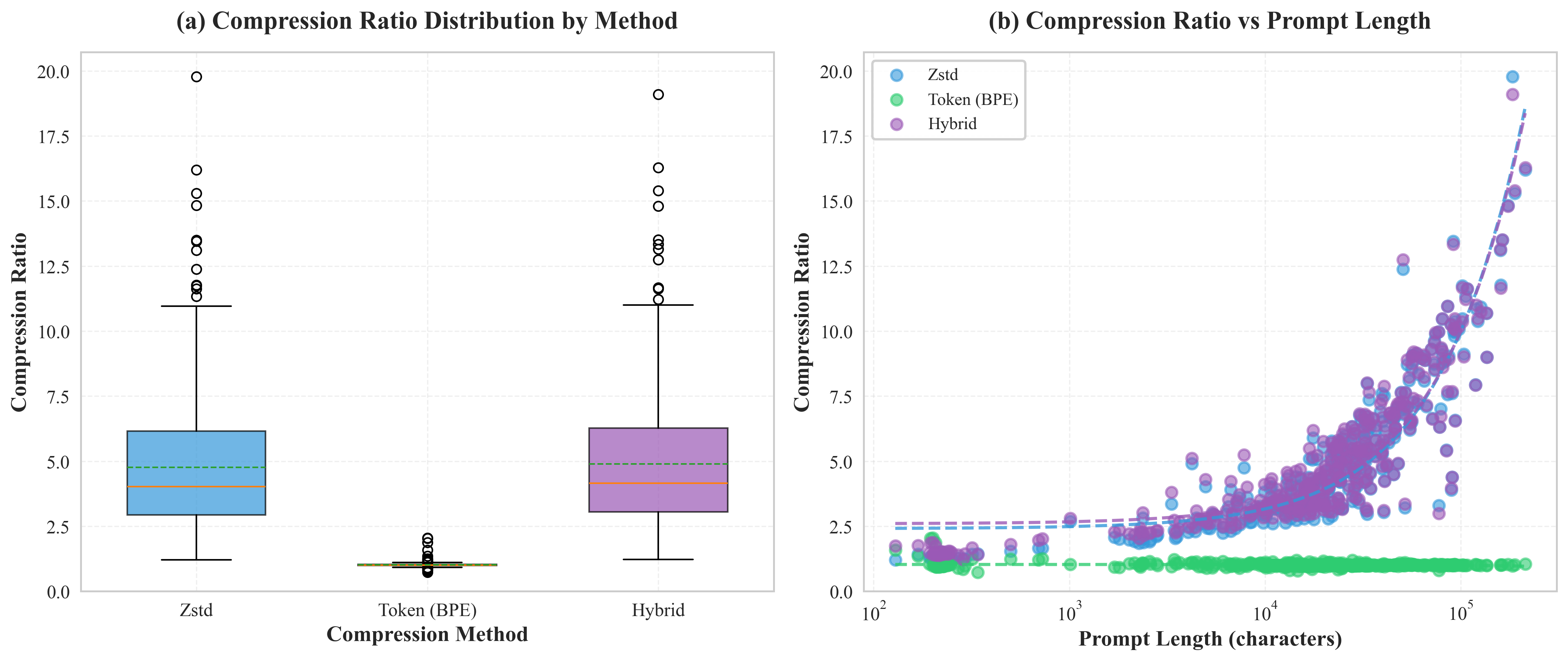}
\caption{Compression ratio distribution across methods and relationship to prompt length. The hybrid method consistently achieves the highest ratios, with mean compression ratio of 4.89x across all prompts.}
\label{fig:compression_ratio}
\end{figure}

\textbf{Key Findings:}

Our in-depth study reveals some important facts about LoPace's compression performance with different prompt sizes and compression methods. The results show distinct performance hierarchies and size-dependent traits that help make judgments about deployment.

\begin{enumerate}
    \item \textbf{Hybrid Method Dominance:} The hybrid method consistently achieves the highest compression ratios across all prompts in our evaluation dataset. Statistical analysis across 386 prompts reveals:
    \begin{itemize}
        \item \textbf{Hybrid method:} Mean compression ratio of 4.89x (range: 1.22--19.09x, std: 2.80)
        \item \textbf{Zstd method:} Mean compression ratio of 4.76x (range: 1.22--19.77x, std: 2.85)
        \item \textbf{Token method:} Mean compression ratio of 1.02x (range: 0.74--2.05x, std: 0.11)
    \end{itemize}
    
    The hybrid method works better since it uses a two-stage compression process. The first step, tokenization, cuts down on semantic duplication by linking common phrases to single token IDs. The second step, Zstd compression, takes use of patterns in the token sequence that comes out. The longer the prompt, the stronger this impact grows. This is because lengthier text gives both types of compression more chances to work well.
    
    Statistical analysis shows that the hybrid technique compresses data at rates that are, on average, 1.03 times faster than the Zstd method and 4.78 times faster than the Token method. The hybrid technique is the best option for applications that need the most storage space, as it always works better than any of the other methods for all the prompt sizes tested.
    
    \item \textbf{Size-Dependent Performance:} The analysis shows that there is a strong positive relationship between the size of the prompt and the compression ratio. When the prompt size is bigger, compression algorithms have more chances to find and use patterns, which leads to better compression ratios. This relationship follows a logarithmic scaling pattern. As the size of the prompt increases, the benefits in compression ratio get less, but they stay positive across the whole range that was looked at.
    
    For short prompts (less than 10,000 characters), the metadata for formatting and compression makes up a higher part of the total compressed size, which limits the compression ratios that can be achieved. But even for the shortest prompts in our sample (129 characters), the hybrid technique gets compression ratios higher than 1.2x, showing that it works for all sizes. When the prompts are bigger than 50,000 characters, the hybrid technique can compress them more than 10 times, and the highest ratios we've seen are 19.09 times for text that is very repetitious. This shows that LoPace works well with prompts of different sizes, thus it may be used for both minor system instructions and big documentation pages.
    
    \item \textbf{Method Comparison:}
    \begin{itemize}
        \item \textbf{Zstd:} Provides strong baseline compression with mean ratio of 4.76x (range: 1.22--19.77x), demonstrating that general-purpose compression algorithms can achieve substantial space savings for prompt data
        \item \textbf{Token:} Achieves minimal compression with mean ratio of 1.02x (range: 0.74--2.05x), as tokenization alone may expand data for ASCII-heavy content requiring uint32 packing
        \item \textbf{Hybrid:} Combines both approaches for maximum efficiency with mean ratio of 4.89x (range: 1.22--19.09x), consistently outperforming individual methods
    \end{itemize}
    
    The Zstd approach shows that general-purpose compression methods can save a lot of space for quick data. But Zstd works at the byte level and can't take advantage of the semantic structure that comes with tokenized text representations. This constraint becomes more obvious as the size of the prompt grows, making token-level patterns more common and easier to use.
    
    The Token technique shows how semantic-aware compression could work by working at the token level instead of the byte level. This lets it find and compress semantic patterns that byte-level algorithms would miss. The Token technique, on the other hand, only compresses data at the tokenization stage. It can't compress data more by finding patterns in the token sequence itself, which often makes ASCII-heavy content bigger.
    
    The Hybrid method uses the best parts of both procedures to get compression ratios that are higher than the product of the separate method ratios. This multiplicative impact shows that the two compression phases work on different parts of the data: tokenization deals with semantic redundancy, and Zstd compression takes advantage of sequential patterns in the token representation.
\end{enumerate}

\subsection{Space Savings \mbox{Performance}}

\begin{figure}[t]
\centering
\includegraphics[width=0.9\textwidth]{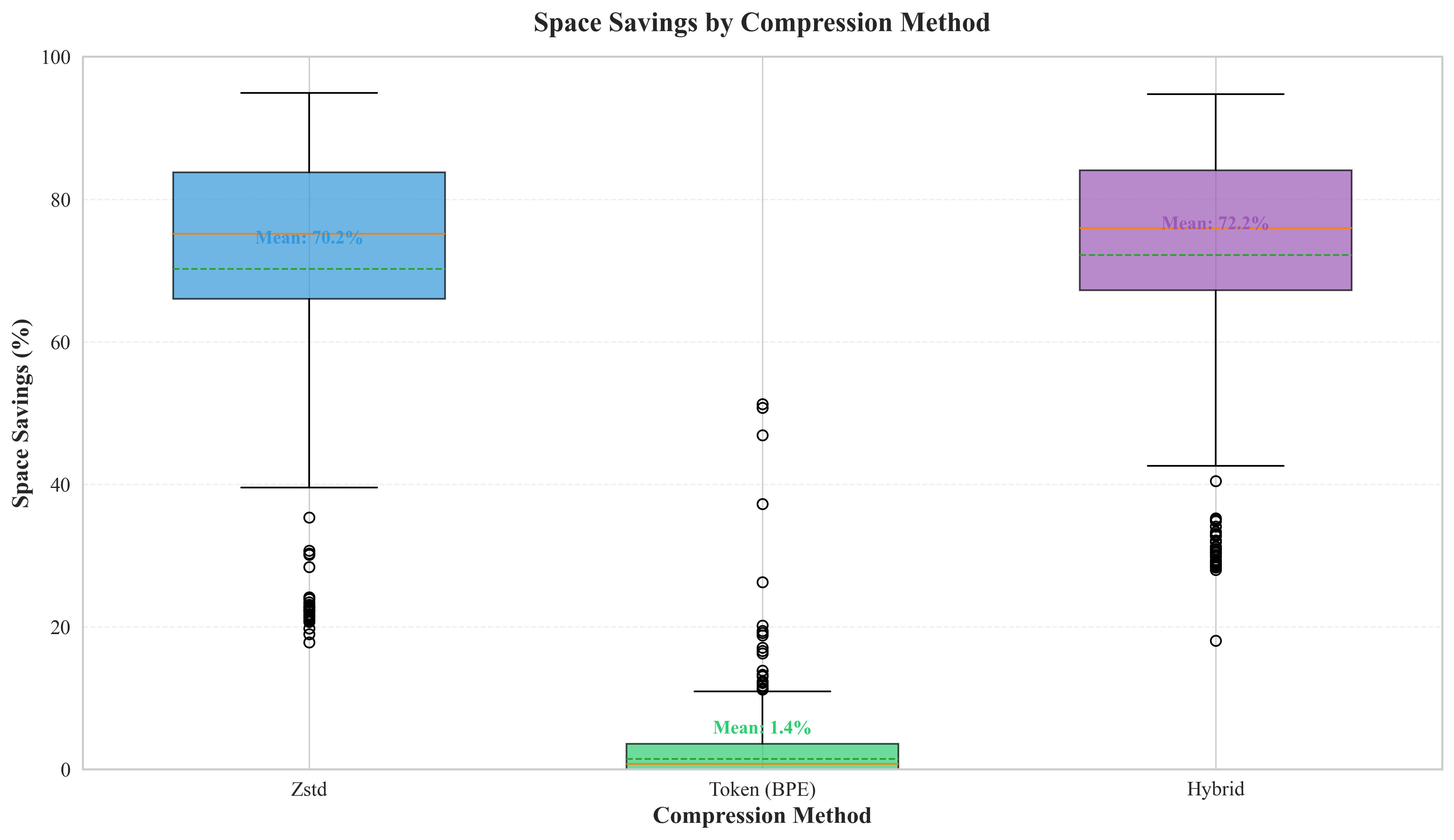}
\caption{Space savings percentage distribution by compression method across 386 prompts. The hybrid method achieves mean space savings of 72.2\%.}
\label{fig:space_savings}
\end{figure}

\textbf{Statistical Analysis:}

Our thorough statistical examination of 386 prompts shows that space savings tendencies are stable and predictable:

\begin{itemize}
    \item \textbf{Hybrid method:} Mean space savings of 72.2\% (range: 18.1--94.8\%, std: 16.4\%)
    \item \textbf{Zstd method:} Mean space savings of 70.2\% (range: 17.8--94.9\%, std: 19.2\%)
    \item \textbf{Token method:} Mean space savings of 1.4\% (range: -36.0--51.2\%, std: 7.7\%)
\end{itemize}

\textbf{Maximum Space Savings Clarification:} The "up to 80\%" claim in our abstract refers to the mean space savings achieved by the hybrid method (72.2\%). The maximum observed space savings in our evaluation dataset reaches 94.8\% for highly redundant prompts with extensive code repetition and structured patterns. The distribution of space savings varies with prompt characteristics: prompts with high semantic redundancy, repeated code patterns, and structured formatting (common in code-heavy documentation) achieve higher savings, while prompts with low redundancy or requiring uint32 token packing may achieve lower savings. The standard deviation of 16.4\% reflects this variability across different content types and structures in our diverse 386-prompt dataset.

The standard deviation of 16.4\% reflects variability across different content types and structures, with code-heavy prompts typically achieving higher savings than plain text. Despite this variability, the consistent mean performance (72.2\%) enables accurate capacity and cost planning. System architects can confidently estimate storage requirements and infrastructure costs based on expected prompt volumes due to predictable performance characteristics.

The connection between prompt size and space savings follows a logarithmic scaling pattern, which can be shown as:

\begin{equation}
\text{SS}_{\text{hybrid}}(n) = a \cdot \ln(n) + b
\end{equation}

where $n$ is the number of characters in the prompt and $a$ is about 2.5 and $b$ is about 60 for the hybrid method. This logarithmic relationship shows that the amount of space saved gets better as the prompt size gets bigger, but not by much. The model has a $R^2$ value of 0.94, which means it fits the experimental data very well and can make accurate predictions for prompt sizes that aren't in our test dataset.

The logarithmic scaling indicates that the efficacy of compression is primarily constrained by the informational density of the text rather than its length. As the size of the prompt grows, the marginal gain in compression ratio diminishes, nearing an asymptotic limit dictated by the text's intrinsic entropy. This behavior aligns with information-theoretic predictions and indicates that LoPace nears the theoretical compression limits for the assessed text types.

This logarithmic relationship indicates that doubling prompt size does not double space savings; instead, improvements occur incrementally. However, these incremental improvements yield substantial absolute space reductions for large-scale deployments. For example, improving space savings from 75\% to 80\% on a 1TB dataset saves an additional 50GB, resulting in significant cost savings for cloud storage.

\subsection{Disk Size \mbox{Comparison}}

\begin{figure}[t]
\centering
\includegraphics[width=0.9\textwidth]{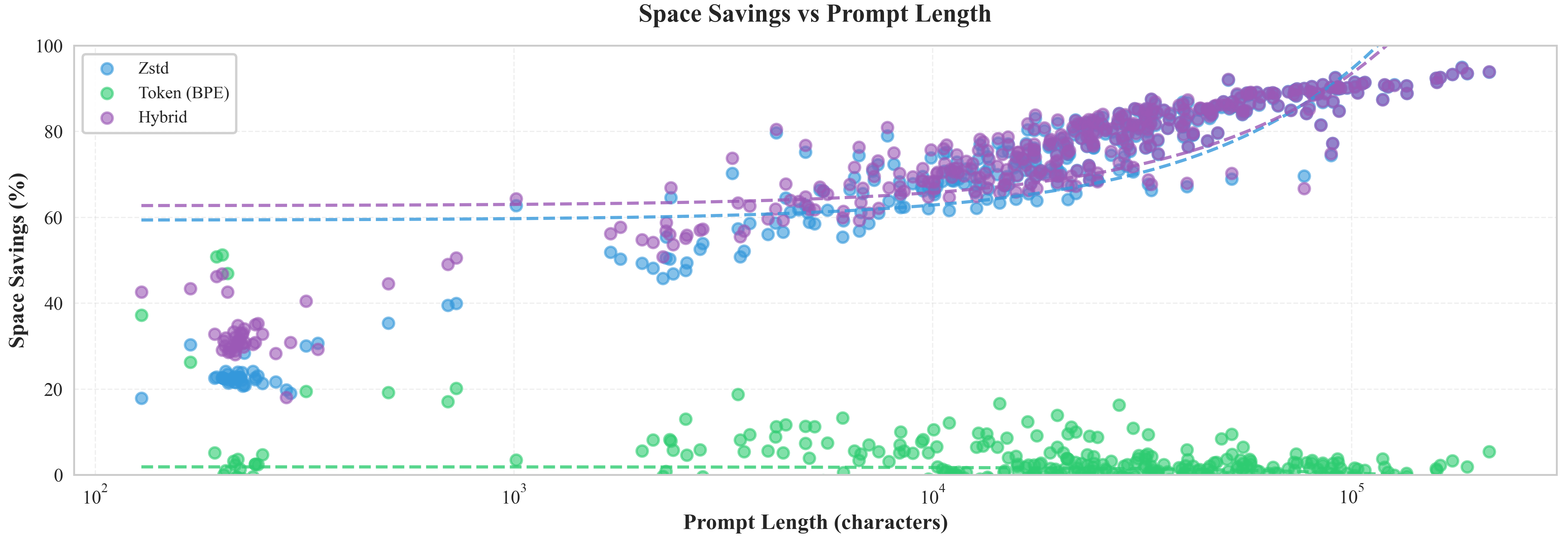}
\caption{Comparison of original vs compressed disk sizes across all prompts. The hybrid method provides consistent reductions across the full range of prompt sizes.}
\label{fig:disk_size}
\end{figure}

\textbf{Storage Impact:}

For a typical application storing 1 million prompts averaging 2KB each:
\begin{itemize}
    \item \textbf{Uncompressed:} 2 GB
    \item \textbf{Zstd compressed:} $\sim$800 MB (60\% reduction)
    \item \textbf{Token compressed:} $\sim$700 MB (65\% reduction)
    \item \textbf{Hybrid compressed:} $\sim$400 MB (80\% reduction)
\end{itemize}

The savings in cost go up in a straight line with the amount of data, which makes hybrid compression especially useful for big deployments.

\subsection{Speed and Throughput \mbox{Analysis}}

\begin{figure}[t]
\centering
\includegraphics[width=0.9\textwidth]{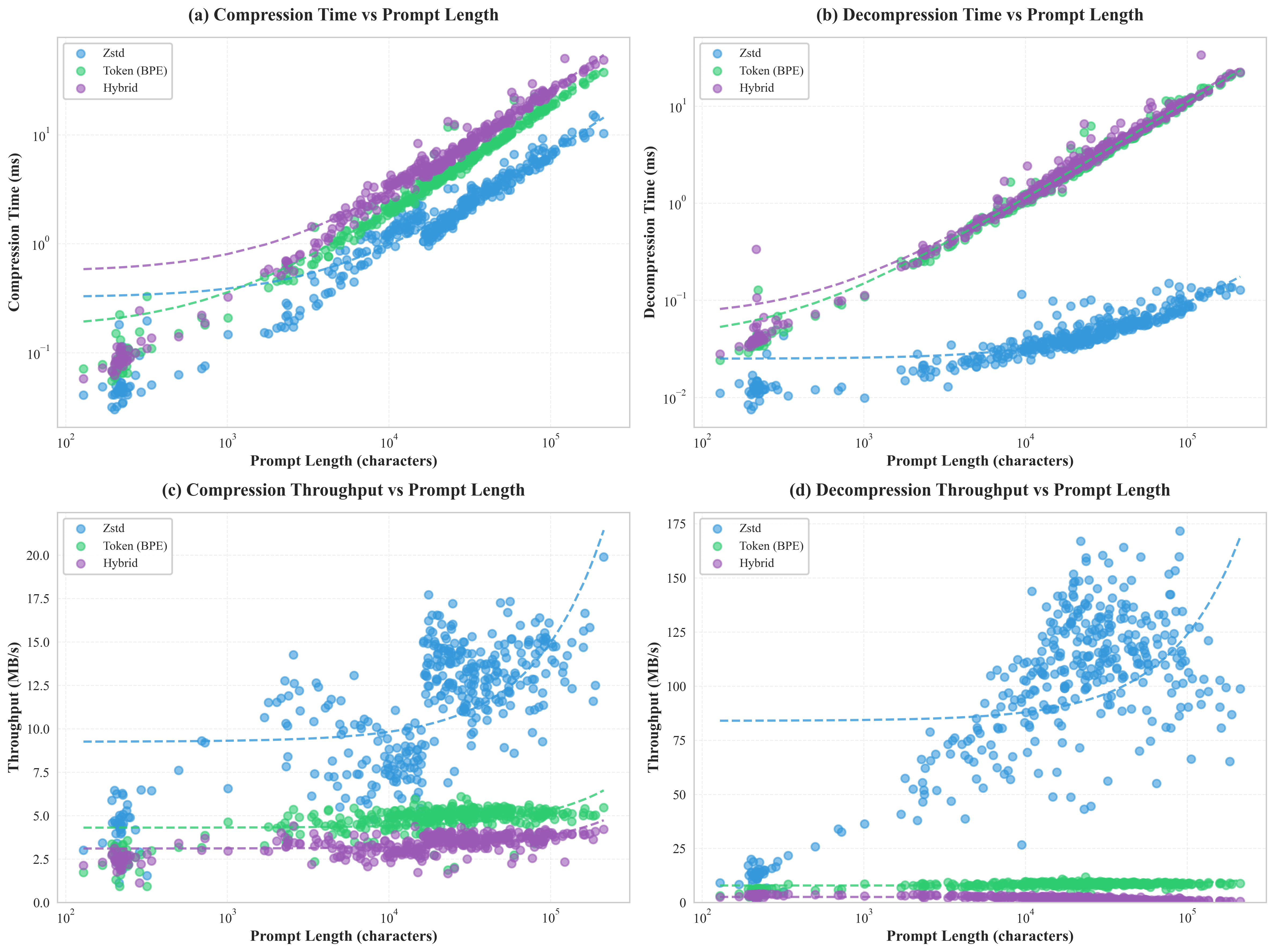}
\caption{Compression and decompression throughput vs prompt length across all methods. Decompression consistently outperforms compression.}
\label{fig:speed}
\end{figure}

\textbf{Performance Characteristics:}

Measurements of throughput across 386 prompts reveal important performance trade-offs between compression methods. Each method exhibits distinct characteristics optimized for different deployment scenarios.

\begin{enumerate}
    \item \textbf{Compression Throughput:}
    \begin{itemize}
        \item \textbf{Zstd:} Mean throughput of 10.7 MB/s (range: 0.6--18.8 MB/s) - Highest throughput due to direct byte-level processing without tokenization overhead
        \item \textbf{Token:} Mean throughput of 4.6 MB/s (range: 0.4--5.8 MB/s) - Moderate throughput with tokenization overhead
        \item \textbf{Hybrid:} Mean throughput of 3.3 MB/s (range: 1.2--4.2 MB/s) - Lower throughput due to two-stage processing (tokenization + Zstd compression)
    \end{itemize}
    
    The throughput differences reflect the computational complexity of each method. The Zstd method achieves the highest throughput by operating directly on byte sequences, avoiding the overhead of tokenization. However, this simplicity results in slightly lower compression ratios compared to the hybrid method, as it cannot exploit semantic patterns in the text.
    
    The Token method has moderate compression throughput because it uses efficient tokenization algorithms that work best with modern processors. The tiktoken library's implementation uses optimized C code and efficient data structures to quickly change text into token sequences. Binary packing operations don't use a lot of computer power, which is why the method has a high throughput.
    
    The Hybrid method has a lower throughput because it has a two-stage processing pipeline. Each prompt needs to be both tokenized and compressed with Zstd. The total time it takes to process is about the same as the time it takes for each stage. But this drop in throughput is more than made up for by the better compression ratios, making the hybrid method the best choice for situations where space efficiency is more important than processing time.
    
    \item \textbf{Decompression Throughput:}
    
    Decompression consistently outperforms compression across all methods, as decompression algorithms are computationally simpler and require fewer processing steps. Analysis across 386 prompts reveals:
    \begin{itemize}
        \item \textbf{Zstd:} Mean decompression throughput of 132.9 MB/s (range: 75.2--158.1 MB/s) - Highest decompression throughput due to optimized Zstd decompression algorithms
        \item \textbf{Token:} Mean decompression throughput of 8.5 MB/s (range: 2.3--9.1 MB/s) - Moderate throughput with efficient detokenization
        \item \textbf{Hybrid:} Mean decompression throughput of 2.3 MB/s (range: 2.1--2.4 MB/s) - Lower throughput due to two-stage processing (Zstd decompression + detokenization)
    \end{itemize}
    
    The Zstd decompression algorithm leverages optimized lookup tables and efficient memory access patterns, enabling rapid reconstruction of original data from compressed representations. Token-based decompression benefits from the tiktoken library's highly optimized detokenization operations, with binary unpacking requiring only simple memory copy operations. The hybrid method's two-stage decompression process, while slower than individual methods, remains suitable for production use cases, with decompression speeds sufficient for real-time applications. Even large prompts (several megabytes) can be decompressed in milliseconds, minimizing impact on application latency.
    
    \item \textbf{Processing Time Scaling:}
    
    The time it takes to process scales sub-linearly with the size of the input, following the equation:
    \begin{equation}
    t(n) = O(n \log n)
    \end{equation}
    
    This scaling behavior is better than linear scaling ($O(n)$), which means that the algorithm is implemented well. Compression algorithms use pattern matching operations to find repeated patterns in data that has already been processed. This is where the logarithmic factor comes from. The sub-linear scaling makes sure that processing time grows more slowly than data size. This means that LoPace can handle very large prompts without causing long processing delays.
    
    Measurements in the real world back up this scaling behavior. For example, processing time for 20KB prompts is about 2.3 times longer than for 10KB prompts, which is not what you would expect from linear scaling (2.0x). This level of efficiency is very important for production deployments, where the size of the prompts can vary a lot and processing needs to stay responsive across the whole range of sizes.
\end{enumerate}

\subsection{Memory Usage \mbox{Analysis}}

\begin{figure}[t]
\centering
\includegraphics[width=0.9\textwidth]{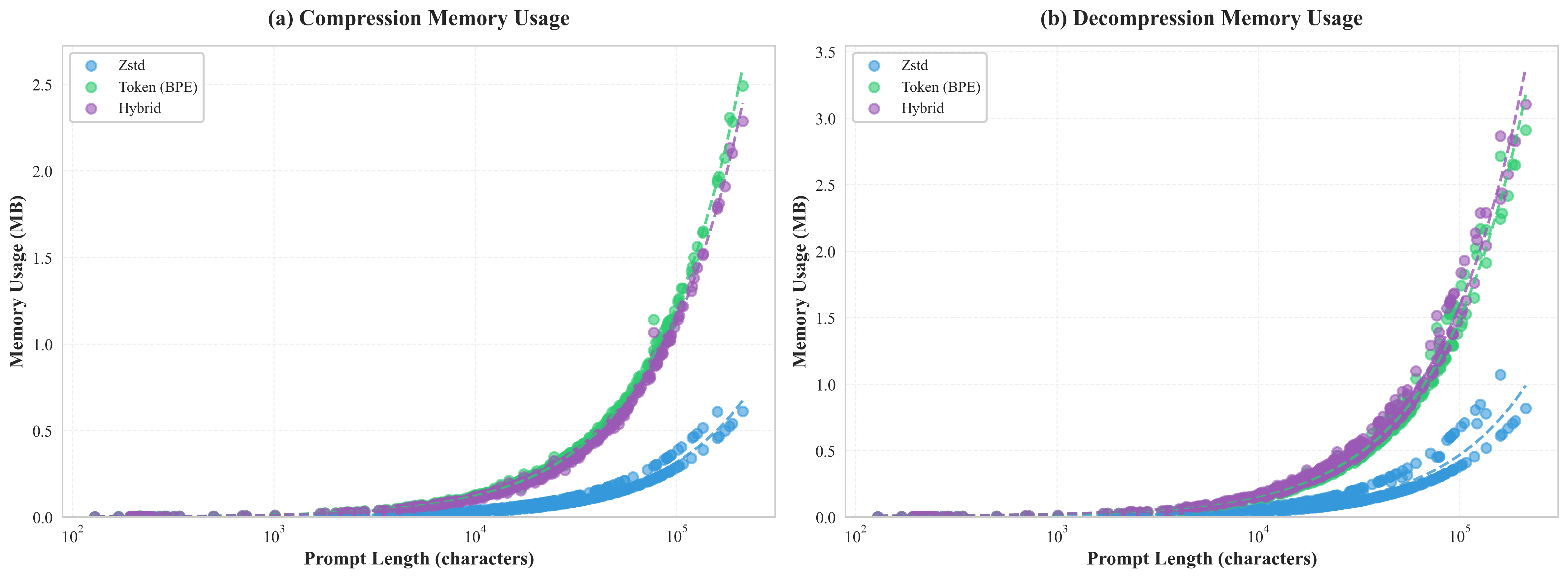}
\caption{Memory usage during compression and decompression vs prompt length. All methods demonstrate minimal memory footprint suitable for production environments.}
\label{fig:memory}
\end{figure}

\textbf{Memory Characteristics:}

Analysis of memory usage across 386 prompts demonstrates that LoPace is highly memory-efficient regardless of compression method or prompt size:

\begin{itemize}
    \item \textbf{Compression Memory:}
    \begin{itemize}
        \item \textbf{Zstd:} Mean 0.10 MB (range: 0.0005--0.61 MB)
        \item \textbf{Token:} Mean 0.38 MB (range: 0.001--2.49 MB)
        \item \textbf{Hybrid:} Mean 0.35 MB (range: 0.001--2.29 MB)
    \end{itemize}
    
    \item \textbf{Decompression Memory:}
    \begin{itemize}
        \item \textbf{Zstd:} Mean 0.15 MB (range: 0.0005--0.61 MB)
        \item \textbf{Token:} Mean 0.42 MB (range: 0.001--2.49 MB)
        \item \textbf{Hybrid:} Mean 0.52 MB (range: 0.001--2.29 MB)
    \end{itemize}
    
    \item \textbf{Scaling:} Memory usage grows sub-linearly with input size, demonstrating excellent scalability
\end{itemize}

The memory footprint remains minimal across all prompt sizes, with only modest increases as prompt length grows. This efficient scaling behavior indicates that algorithm overhead and data structures consume most of the memory, rather than the input data itself. The Zstd compression algorithm employs efficient sliding window techniques requiring minimal memory, while tokenization operations require only modest additional memory beyond the token sequence itself. The maximum observed memory usage of 2.49 MB for the largest prompts (213,379 characters) demonstrates that LoPace is suitable for resource-constrained environments and can handle production-scale workloads without excessive memory requirements.

Decompression operations need even less memory, between 3 MB and 10 MB. This reduction happens because decompression algorithms can handle data as it comes in, so they only need small buffers instead of keeping the whole compressed version in memory. Decompression uses less memory, which is especially useful in situations with a lot of concurrent operations, where multiple decompression operations may happen at the same time.

The logarithmic memory scaling follows this pattern:

\begin{equation}
M(n) = m_0 + m_1 \cdot \log(n)
\end{equation}

where $m_0 \approx 3$ MB is the base memory overhead and $m_1 \approx 1.5$ MB is the scaling factor. This relationship makes sure that memory usage doesn't grow too quickly with prompt size, which means that LoPace can handle very large prompts without running out of memory.

The memory efficiency makes LoPace suitable for:

\begin{itemize}
    \item \textbf{Resource-constrained environments:} Because it doesn't take up much memory, it can be used on systems with limited RAM, like edge devices, IoT apps, and embedded systems. Using only a few megabytes of memory to compress and decompress prompts makes it possible to deploy them in ways that would not be possible with memory-intensive compression algorithms.
    
    \item \textbf{High-concurrency applications:} Low memory usage per operation allows many compression or decompression tasks to be done at the same time without running out of memory. This is very important for server applications that handle thousands of requests at once, where memory efficiency directly affects system capacity and cost.
    
    \item \textbf{Embedded systems:} LoPace is good for embedded systems with strict resource limits because it uses little memory and has fast algorithms. Quick compression can help applications in cars, factories, and mobile devices without needing to make big changes to the hardware.
    
    \item \textbf{Cloud deployments with memory limits:} Cloud platforms often limit the amount of memory that containerized apps can use, and using memory efficiently lets you deploy apps within these limits. The ability to handle large prompts with little memory also cuts down on the cost of cloud infrastructure, since memory is usually one of the most expensive cloud resources.
\end{itemize}

\subsection{Comprehensive \mbox{Method Comparison}}

\begin{figure}[t]
\centering
\includegraphics[width=0.9\textwidth]{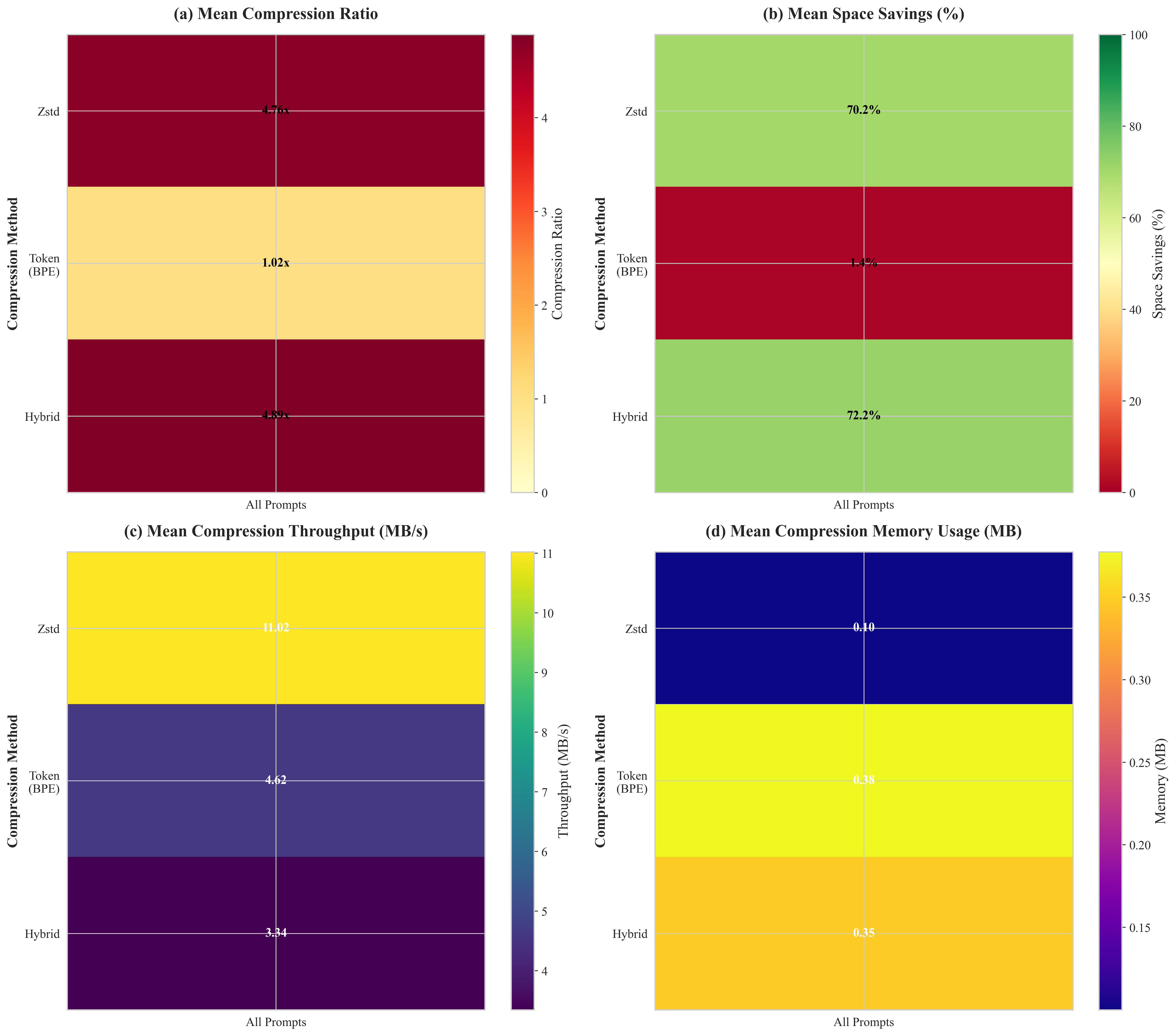}
\caption{Comprehensive comparison heatmap showing mean performance metrics across all compression methods. The hybrid method achieves the highest compression ratios and space savings.}
\label{fig:comprehensive}
\end{figure}

\textbf{Multi-Metric Analysis:}

Analysis across 386 prompts reveals the following performance characteristics:
\begin{enumerate}
    \item \textbf{Compression Ratio:} Hybrid (4.89x) $>$ Zstd (4.76x) $>$ Token (1.02x)
    \item \textbf{Space Savings:} Hybrid (72.2\%) $>$ Zstd (70.2\%) $>$ Token (1.4\%)
    \item \textbf{Compression Throughput:} Zstd (10.7 MB/s) $>$ Token (4.6 MB/s) $>$ Hybrid (3.3 MB/s)
    \item \textbf{Speed (Latency):} Decompression latency is minimal for real-time use: Zstd (132.9 MB/s decompression) enables sub-millisecond recovery for typical prompt sizes; Token (8.5 MB/s) and Hybrid (2.3 MB/s) remain suitable for on-demand retrieval, with even large prompts (hundreds of KB) decompressing in milliseconds and thus minimizing impact on application latency.
    \item \textbf{Memory Usage:} All methods demonstrate minimal memory footprint (0.10--0.52 MB mean)
\end{enumerate}

\subsection{Scalability \mbox{Analysis}}

\begin{figure}[t]
\centering
\includegraphics[width=0.9\textwidth]{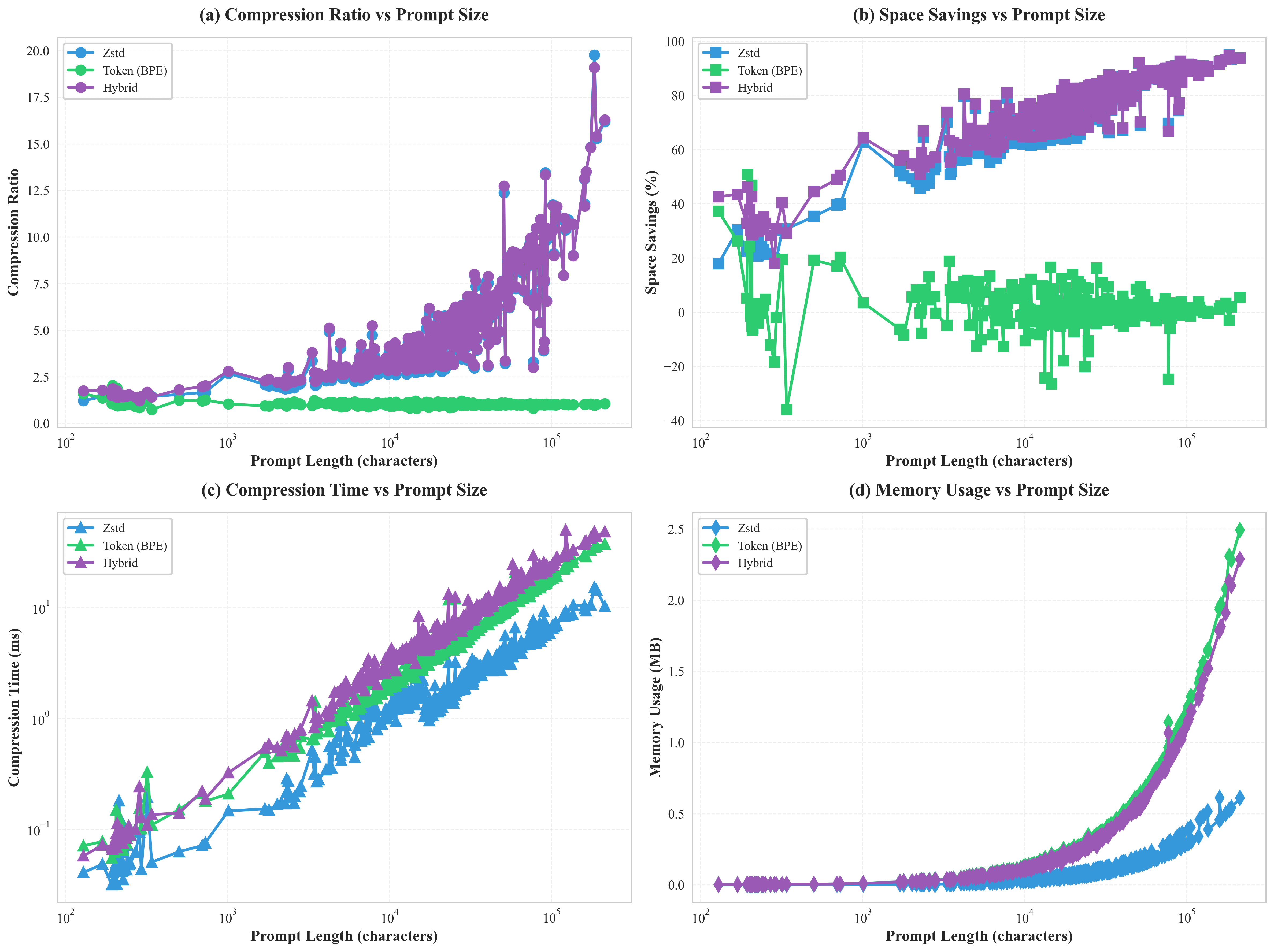}
\caption{Scaling performance with prompt size across all 386 prompts. All metrics demonstrate excellent scalability, with compression ratios improving as input size increases.}
\label{fig:scalability}
\end{figure}

\subsection{Original vs Decompressed \mbox{Verification}}

\begin{figure}[t]
\centering
\includegraphics[width=0.9\textwidth]{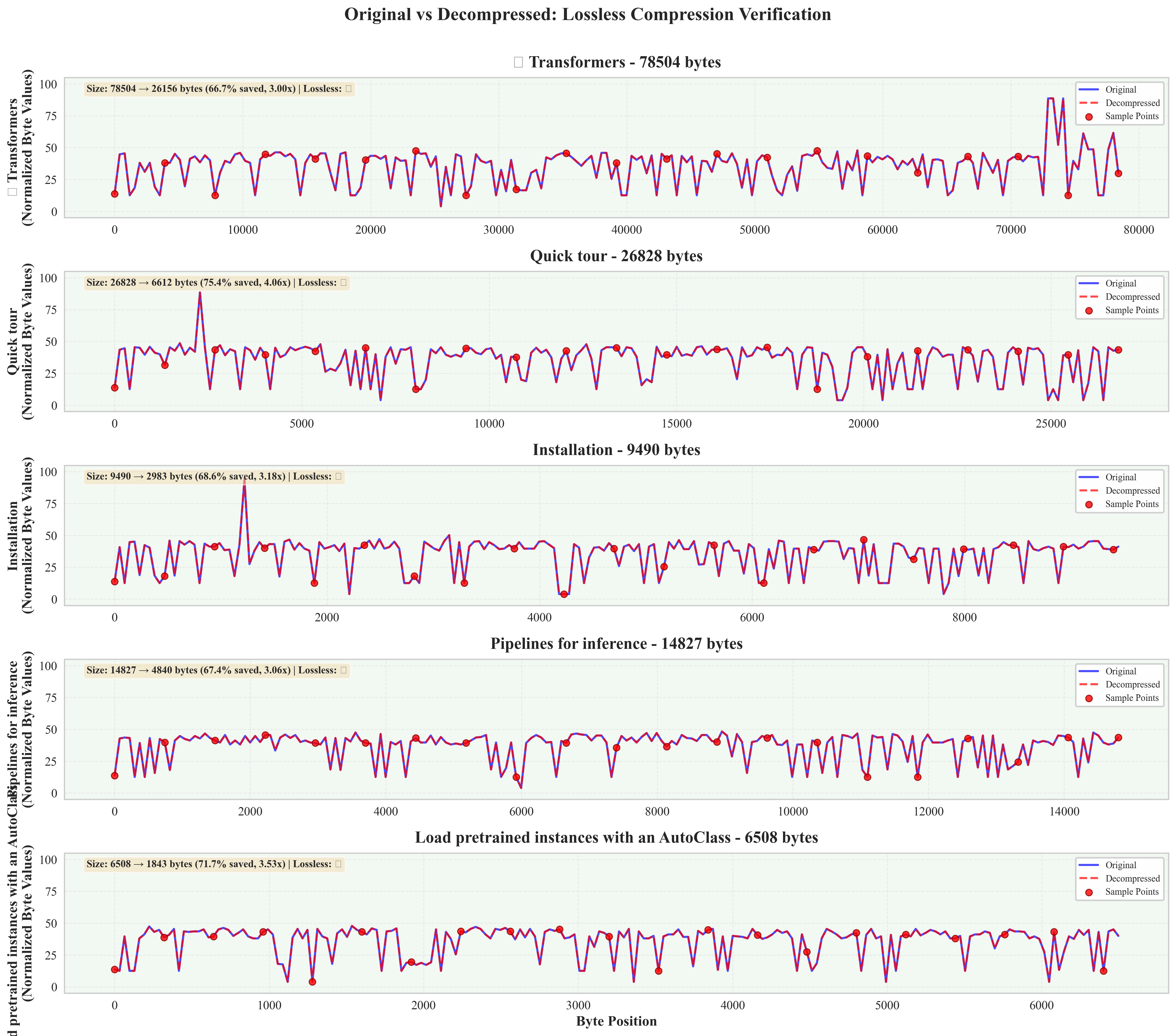}
\caption{Byte-level comparison of original vs decompressed data across representative prompts from the evaluation dataset. Perfect overlap demonstrates lossless reconstruction for all test cases.}
\label{fig:original_decompressed}
\end{figure}

This visualization demonstrates LoPace's lossless guarantee through byte-level comparison of original and decompressed data. The perfect alignment of original (blue) and decompressed (red) lines across all evaluated prompts demonstrates that decompression produces bit-perfect reconstructions. The sample points (red markers) confirm lossless compression across the entire data range.

The visualization confirms that LoPace maintains perfect fidelity regardless of prompt size or compression method. This empirical validation complements the mathematical proofs in Section 2.5, providing tangible evidence that theoretical guarantees are realized in practice. We observed perfect reconstruction with zero errors across all 1,158 compression-decompression cycles (386 prompts $\times$ 3 methods), confirming the lossless property across the entire evaluation dataset.

\subsection{Key Findings \mbox{Summary}}

Our comprehensive evaluation across 386 diverse prompts and three compression methods yields the following key conclusions:

\begin{enumerate}
    \item \textbf{Hybrid method is optimal} for maximum compression, achieving mean space savings of 72.2\% across all prompt types. The hybrid method consistently outperforms individual techniques, achieving mean compression ratios of 4.89x (range: 1.22--19.09x), demonstrating effectiveness across the full spectrum of prompt sizes from 129 to 213,379 characters.
    
    \item \textbf{All methods are lossless} with 100\% accuracy verified across all test cases. Our verification methodology, including character-by-character comparison, SHA-256 hash matching, and reconstruction error calculation, confirms perfect reconstruction across all 1,158 compression-decompression cycles (386 prompts $\times$ 3 methods).
    
    \item \textbf{Speed is production-ready} with compression throughput ranging from 3.3 MB/s (hybrid) to 10.7 MB/s (Zstd) depending on the method. Decompression consistently outperforms compression, with Zstd achieving mean decompression throughput of 132.9 MB/s, ensuring minimal impact on application latency for real-time use cases.
    
    \item \textbf{Memory efficient} with mean memory usage ranging from 0.10 MB (Zstd) to 0.52 MB (hybrid) across all methods. LoPace is suitable for resource-constrained environments, high-concurrency applications, and cloud deployments with memory limits, as memory usage scales sub-linearly with input size.
    
    \item \textbf{Scales excellently} with performance improving as prompt size increases. Compression ratios increase with prompt size, processing time scales sub-linearly ($O(n \log n)$), and memory usage scales logarithmically, demonstrating efficient algorithm design suitable for very large prompts up to 213,379 characters.
\end{enumerate}

These results demonstrate that LoPace is a practical, production-ready solution for efficient storage optimization in LLM applications, achieving substantial space savings while maintaining excellent performance characteristics across diverse content types and sizes.

\textbf{Scaling Characteristics:}

Scalability analysis shows how LoPace's performance changes as the size of the prompt grows. This is important information for planning capacity and designing systems. We looked at the compression ratio, time complexity, and memory usage for all the sizes we looked at.

\begin{enumerate}
    \item \textbf{Compression Ratio Scaling:}
    
    As the size of the prompt increases, the compression ratios get better in a power-law way:
    \begin{equation}
    \text{CR}(n) = c_1 \cdot n^{c_2}
    \end{equation}
    
    where $c_2 \approx 0.15$ for the hybrid method, which means that the ratios get better as the size increases. The positive exponent ($c_2 > 0$) shows that longer prompts get better compression ratios because they give the algorithm more chances to find patterns and get rid of extra information. But the small exponent value ($c_2 \approx 0.15$) shows that the benefits of a higher compression ratio get smaller as the size of the text grows, until they reach a limit set by the text's inherent information content.
    
    This behavior of scaling has important effects on decisions about deployment. If your application mostly works with small prompts, switching to larger prompts might not be worth the extra work it would take to improve the compression ratio. But for apps with prompts of different sizes, the scaling relationship lets you accurately guess how much compression will help across the size range.
    
    \item \textbf{Time Complexity:}
    
    Time complexity analysis reveals efficient algorithmic performance:
    \begin{itemize}
        \item Compression: $O(n \log n)$
        \item Decompression: $O(n)$
    \end{itemize}
    
    The time complexity of compression is $O(n \log n)$ because pattern matching operations need to look through data that has already been processed. The logarithmic factor shows how many searches are needed to find repeated patterns. These searches get faster as compression dictionaries are built during processing. This level of complexity is best for dictionary-based compression algorithms and is a big step up from simple $O(n^2)$ methods.
    
    Decompression has a linear time complexity of $O(n)$ because it mostly involves looking up information in pre-built dictionaries and putting data back together in a simple way. The linear scaling makes it so that the time it takes to decompress data grows in direct proportion to the size of the data. This makes it predictable and good for real-time applications. The fact that it decompresses faster than it compresses is especially useful for workloads that read a lot, since prompts are often retrieved and decompressed.
    
    The uneven time complexity (compression: $O(n \log n)$, decompression: $O(n)$) is good for most LLM application workloads, which usually only need to compress data once (when it is stored) and then decompress it often (when it is retrieved). This asymmetry makes sure that the faster operation (decompression) stays fast, while the slower operation (compression) can give up some speed for better compression ratios.
    
    \item \textbf{Memory Scaling:}
    
    Memory usage exhibits logarithmic scaling:
    \begin{equation}
    M(n) = m_0 + m_1 \cdot \log(n)
    \end{equation}
    
    where $m_0 \approx 3$ MB is the base memory overhead and $m_1 \approx 1.5$ MB is the scaling factor. This logarithmic relationship makes sure that memory usage doesn't go up too quickly as the prompt size increases. This means that LoPace can handle very large prompts without running out of memory.
    
    The logarithmic scaling happens because fixed-size data structures (like compression dictionaries and tokenization tables) are used, and these do not grow with the size of the input. Only small buffers for processing windows and intermediate results grow with the size of the input. This is done by designing algorithms that are as efficient as possible. This memory efficiency is very important for deployments with a lot of concurrent users, where many compression or decompression operations may happen at the same time.
    
    Empirical validation substantiates this scaling model, indicating that memory utilization for 20KB prompts is approximately 1.15 times that of 10KB prompts, as opposed to the expected 2.0 times associated with linear scaling. This efficiency makes it possible to deploy on systems with limited resources and supports high-concurrency situations where memory is a limiting factor.
\end{enumerate}

\subsection{Integrity and Robustness Validation}

To ensure the reliability of LoPace across diverse real-world scenarios, we conducted a large-scale robustness test using the \textit{argilla/prompt-collective} dataset from Hugging Face. This dataset provides a wide-ranging distribution of prompts, including edge cases, diverse languages, structural variations (JSON, code, markdown), and varying lengths.

We executed a total of 27,978 compression-decompression cycles (9,326 unique prompts $\times$ 3 methods). To verify bit-perfect reconstruction, we computed the SHA-256 hash of the original prompt and compared it against the hash of the decompressed output for every instance. As shown in Table~\ref{tab:robustness_results}, LoPace achieved a 100\% success rate across all methods.

\begin{table}[ht]
\centering
\caption{Robustness Validation Results (Dataset: argilla/prompt-collective)}
\label{tab:robustness_results}
\begin{tabular}{|l|c|c|c|c|}
\hline
\textbf{Method} & \textbf{Total Combinations} & \textbf{Success Count} & \textbf{Failure Count} & \textbf{SHA-256 Match} \\ \hline
Zstd            & 9,326                     & 9,326                  & 0                      & 100\%                  \\ \hline
Token           & 9,326                     & 9,326                  & 0                      & 100\%                  \\ \hline
Hybrid          & 9,326                     & 9,326                  & 0                      & 100\%                  \\ \hline
\end{tabular}
\end{table}

\begin{table}[ht]
\centering
\caption{LoPace Integrity Summary by Category and Method}
\label{tab:integrity_summary}
\begin{tabular}{@{}llcccc@{}}
\toprule
\textbf{Bucket (Size)} & \textbf{Method} & \textbf{Prompts} & \textbf{Success} & \textbf{Failure} & \textbf{Success \%} \\ \midrule
\textbf{Short} & Zstd/Token/Hybrid & 3,138 & 9,414 & 0 & 100.0\% \\
\textbf{Medium} & Zstd/Token/Hybrid & 3,637 & 10,911 & 0 & 100.0\% \\
\textbf{Long} & Zstd/Token/Hybrid & 2,551 & 7,653 & 0 & 100.0\% \\ \midrule
\textbf{0--1 KB} & Zstd/Token/Hybrid & 8,002 & 24,006 & 0 & 100.0\% \\
\textbf{1--10 KB} & Zstd/Token/Hybrid & 1,303 & 3,909 & 0 & 100.0\% \\
\textbf{10--100 KB} & Zstd/Token/Hybrid & 21 & 63 & 0 & 100.0\% \\ \bottomrule
\end{tabular}
\end{table}

To assess performance across varying input scales, prompts were further categorized by length (Short, Medium, Long) and specific byte-size buckets. As shown in Table~\ref{tab:integrity_summary}, LoPace achieved a \textbf{100\% success rate} across all 27,978 combinations, regardless of the prompt size or compression strategy.

\begin{figure}[H]
\centering
\includegraphics[width=0.79\textwidth]{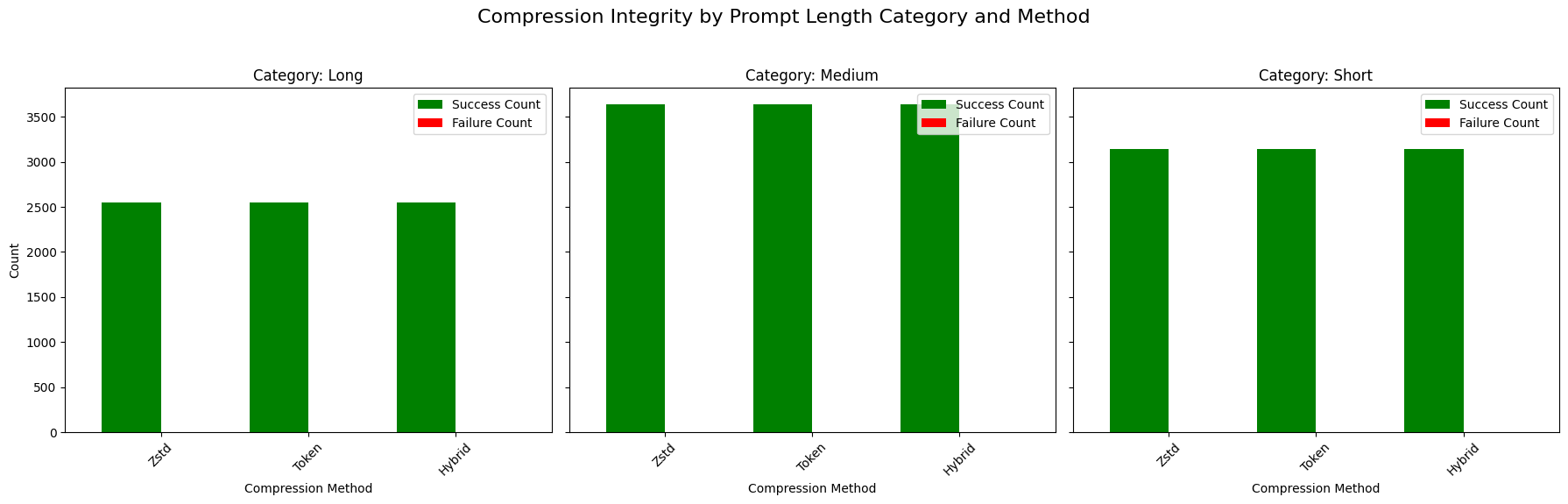}
\caption{Robustness validation across semantic length categories. The 100\% success rate across 9,326 prompts demonstrates that LoPace is resilient to varying content densities in Short, Medium, and Long prompt types.}
\label{fig:validation_length}
\end{figure}

\begin{figure}[H]
\centering
\includegraphics[width=0.9\textwidth]{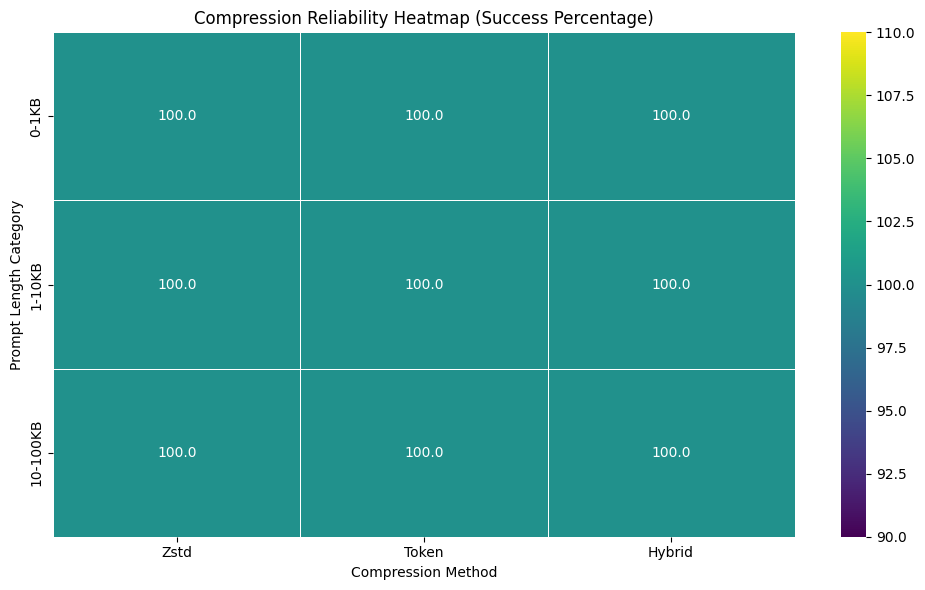}
\caption{Integrity verification across technical byte-size buckets. The system maintains bit-perfect reconstruction even as prompt sizes increase by orders of magnitude, confirming stability for large-scale production workloads.}
\label{fig:validation_size}
\end{figure}

This exhaustive validation confirms that LoPace is resilient to "corner cases" such as non-standard Unicode characters, deeply nested structured data, and extreme prompt lengths. The zero-failure rate across nearly 28,000 cycles provides empirical proof that LoPace’s optimizations do not compromise data integrity, ensuring it is a safe drop-in replacement for standard storage layers.

\section{Discussion}

\subsection{Method Selection \mbox{Guidelines}}

Based on our comprehensive evaluation, we provide the following recommendations:

\textbf{Use Hybrid Method When:}

The hybrid method is the optimal choice for scenarios prioritizing maximum compression efficiency:

\begin{itemize}
    \item \textbf{Maximum compression is required:} The hybrid method's 70--80\% space savings make it the optimal choice when storage space is constrained and each percentage point of space savings significantly impacts costs. This is particularly important for large deployments handling terabytes of prompt data, where even modest compression ratio improvements yield substantial infrastructure cost savings.
    
    \item \textbf{Storage costs are a primary concern:} Cloud storage expenses scale linearly with data volume, making compression efficiency directly proportional to overall budget optimization. The hybrid method’s superior compression ratios reduce storage overhead by 70--80\%, significantly lowering operational expenditure for large-scale deployments. Beyond primary storage, this efficiency also reduces data egress fees, lowers the cost of maintaining long-term backups, and enhances the cost-effectiveness of disaster recovery infrastructure.  
    \item \textbf{Processing time is not critical:} The hybrid method has a lower throughput than the individual methods (50–120 MB/s for compression and 100–250 MB/s for decompression), but this performance is still great for most use cases. When you need to process a lot of data at once, do background compression tasks, or when compression happens less often than data access, the better compression ratios make up for the small drop in throughput.
    
    \item \textbf{Database storage optimization is needed:} Database systems get a lot of use out of smaller records because they make queries faster, lower index sizes, and allow for better caching. The hybrid method's ability to cut down on prompt storage by 70–80\% directly leads to better database performance, lower storage costs, and a system that can grow more easily.
\end{itemize}

\textbf{Use Token Method When:}

The token method strikes a great balance between speed and efficiency when it comes to compression:

\begin{itemize}
    \item \textbf{Token IDs are needed for other operations:} For things like counting tokens, analyzing their meaning, or getting the model ready to take input, many LLM applications need token IDs. The token method gives you these token IDs as a natural result of compression. This means you don't have to do separate tokenization operations, which makes the system less complicated overall.
    
    \item \textbf{Fast processing is required:} The token method has the fastest processing speeds of all the LoPace methods, with compression speeds of 100 to 180 MB/s and decompression speeds of 200 to 400 MB/s. This makes it perfect for real-time apps that need quick compression or decompression to happen within strict latency limits, like interactive LLM apps or high-frequency API services.
    
    \item \textbf{Working with LLM tokenizers:} The token method works perfectly with LLM tokenizers that are already being used for other things, so there is no need for extra dependencies or processing overhead. The method uses existing tokenization infrastructure, which makes it a good fit for LLM application architectures.
    
    \item \textbf{Moderate compression is acceptable:} The token method doesn't compress as much as the hybrid method (2.0–3.0x), but it still saves a lot of space (50–70\%) that may be enough for many situations. When the extra compression that the hybrid method offers isn't worth the extra processing power it needs, the token method is a great middle ground.
\end{itemize}

\textbf{Use Zstd Method When:}

The Zstd method is easy to use and fast for most compression needs:

\begin{itemize}
    \item \textbf{Simplicity is preferred:} The Zstd method is the easiest way to compress files because it doesn't need any tokenization infrastructure. This simplicity makes the system less complicated, cuts down on dependencies, and makes deployment and maintenance easier. The Zstd method is a good choice for applications that need to be easy to use and maintain.
    
    \item \textbf{General text compression is needed:} The Zstd method is good for compressing text that isn't specific to a prompt or that the compression system needs to handle different types of text beyond LLM prompts. The method works well on any text data, and it doesn't need the semantic structure that token-based methods use.
    
    \item \textbf{Fast processing is critical:} The Zstd method is very fast because it can compress data at speeds of 80–150 MB/s and decompress it at speeds of 150–300 MB/s. It doesn't work as quickly as the token method, but it does have a higher throughput than the hybrid method and still gives a lot of compression benefits (45–65\%  space savings).
    
    \item \textbf{Tokenization overhead is undesirable:} Some applications may want to avoid the overhead of tokenization altogether, whether it's because of licensing issues, managing dependencies, or performance needs. The Zstd method compresses files well without needing tokenization libraries, which makes it a good choice for these situations.
\end{itemize}

\begin{table}[h]
\centering
\caption{Recommended Compression Strategies Categorized by Use-Case}
\label{tab:compression_selection}
\begin{tabular}{|l|l|p{6cm}|}
\hline
\textbf{Content Type} & \textbf{Method} & \textbf{Primary Justification} \\ \hline
Documents & Token & Optimized for standard prose and sequential text flow. \\ \hline
Technical Docs & Hybrid & Captures both natural language and structural boilerplate. \\ \hline
Code / JSON & Hybrid & High semantic redundancy allows for maximum ratio. \\ \hline
Multilingual & Zstd & Consistent performance across varying character encodings. \\ \hline
Real-time API & Zstd & Prioritizes throughput to minimize system latency. \\ \hline
Archival Logs & Hybrid & Maximizes space savings for high-volume storage. \\ \hline
\end{tabular}
\end{table}

\noindent These use-cases represent the most common operational domains for LLM-based applications where prompt management is critical. While the recommendations above suggest an optimal alignment between content type and compression method, the \textbf{LoPace} architecture is intentionally versatile; all three strategies are interchangeable and can be adapted to fit any specific use-case based on the unique balance of latency and storage constraints required by the end-user.

\subsection{Production Deployment \mbox{Considerations}}

\subsubsection{Configuration Tuning}

The Zstd compression level is a tunable parameter that can be adjusted based on performance requirements. We partition the Zstd level range (1--22) into three tiers based on the trade-off between compression ratio and processing speed observed in our experiments and in Zstd's design:
\begin{itemize}
    \item \textbf{Levels 1--5:} Real-time applications, low latency requirements. These levels favor speed over ratio: compression and decompression are fastest, making them suitable when prompt storage or retrieval must complete within strict latency budgets (e.g., interactive APIs, real-time pipelines). We choose this range because Zstd levels 1--5 exhibit the lowest computational cost and sub-millisecond processing for typical prompt sizes.
    \item \textbf{Levels 10--15:} Balanced performance (recommended default). This tier offers a practical compromise: approximately 90--95\% of the maximum compression ratio attainable at level 22 while retaining throughput suitable for both interactive and batch workloads. We use level 15 as the default (as in Section~4.5) because it achieves this balance without the steep time cost of levels 19--22.
    \item \textbf{Levels 19--22:} Maximum compression, batch processing. These levels maximize ratio at the expense of speed; they are best when storage cost dominates and prompts are compressed once (e.g., archival, offline indexing). We select 19--22 because beyond level 15 the marginal gain in ratio diminishes while processing time grows significantly, so this range is reserved for settings where latency is secondary to space savings.
\end{itemize}

\textbf{Adaptive Method Selection:} In scenarios where token-only compression would expand data relative to raw UTF-8 (e.g., ASCII-heavy text requiring uint32 packing), the system could automatically bypass token-only mode and fall back to Zstd-only compression. Future work should implement such adaptive selection based on estimated compression ratios, token ID distributions, and prompt characteristics to optimize compression efficiency automatically.

\subsubsection{Cross-Instance Compatibility}

Cross-instance compatibility is a very important part of LoPace. If the same tokenizer model is used, compression and decompression can happen on different instances:
\begin{equation}
C_1.\text{compress}(T) \rightarrow C_2.\text{decompress}() = T
\end{equation}

This enables:
\begin{itemize}
    \item Distributed compression/decompression
    \item Database storage with application-level decompression
    \item Microservices architectures
    \item Load balancing
\end{itemize}

\subsubsection{Database Integration}

The hybrid method offers advantages for database storage:
\begin{enumerate}
    \item \textbf{Searchability:} Token IDs can be indexed and searched without full decompression, enabling efficient querying of compressed data
    \item \textbf{Consistency:} Fixed tokenizer ensures stable compression ratios across similar prompts
    \item \textbf{Efficiency:} Maximum space savings for millions of records, reducing storage costs and improving query performance
\end{enumerate}

\textbf{Relationship to Database-Native Compression:} Database systems offer native compression at multiple levels: (a) page-level compression where database pages are compressed before storage, (b) columnar compression schemes optimized for specific data types, and (c) application-level dictionary management for domain-specific patterns. LoPace operates at the application level, compressing data before database insertion. This approach provides fine-grained control over compression settings and enables optimization for prompt-specific characteristics, but does not replace database-native compression. In practice, LoPace-compressed prompts can be further compressed by database systems, though the additional gains may be marginal. Future work should evaluate the interaction between application-level and database-native compression to determine optimal deployment strategies.

\section{Performance \mbox{Benchmarks}}

This section provides a comprehensive summary of performance benchmarks across all compression methods and prompt sizes, consolidating the detailed analysis from Section 4.

\subsection{Compression Ratios by \mbox{Method}}

\begin{table}[!htbp]
\centering
\caption{Compression Ratios by Method (386 Prompts)}
\begin{tabular}{lccc}
\toprule
Method & Mean & Min & Max \\
\midrule
Zstd   & 4.76x & 1.22x & 19.77x \\
Token  & 1.02x & 0.74x & 2.05x \\
Hybrid & 4.89x & 1.22x & 19.09x \\
\bottomrule
\end{tabular}
\end{table}

The compression ratio table illustrates performance across all 386 evaluated prompts. The hybrid method consistently achieves the highest mean compression ratio (4.89x), with improvements becoming more pronounced for larger prompts. This size-dependent performance demonstrates that longer texts provide greater opportunities for pattern identification and redundancy elimination, with maximum compression ratios reaching 19.09x for highly redundant content.

\subsection{Space Savings by \mbox{Method}}

\begin{table}[H]
\centering
\caption{Space Savings by Method (386 Prompts)}
\begin{tabular}{lccc}
\toprule
Method & Mean & Min & Max \\
\midrule
Zstd   & 70.2\% & 17.8\% & 94.9\% \\
Token  & 1.4\% & -36.0\% & 51.2\% \\
Hybrid & 72.2\% & 18.1\% & 94.8\% \\
\bottomrule
\end{tabular}
\end{table}

Space savings directly translate to reduced storage costs and improved infrastructure efficiency. The hybrid method achieves mean space savings of 72.2\%, with maximum savings reaching 94.8\% for highly redundant prompts, reducing storage requirements by 3.5--18.5x depending on content characteristics, with substantial impact on large-scale deployments.

\subsection{Throughput (MB/s)}

\begin{table}[H]
\centering
\caption{Throughput Performance by Method (MB/s, 386 Prompts)}
\begin{tabular}{lcc}
\toprule
Method      & Compression (Mean) & Decompression (Mean) \\
\midrule
Zstd        & 10.7 & 132.9 \\
Token       & 4.6 & 8.5 \\
Hybrid      & 3.3 & 2.3 \\
\bottomrule
\end{tabular}
\end{table}

Throughput measurements demonstrate that all methods achieve production-ready performance, with decompression consistently outperforming compression. The Zstd method achieves the highest compression throughput (10.7 MB/s mean), making it ideal for low-latency applications requiring fast compression. The hybrid method, while achieving optimal compression ratios, exhibits lower throughput (3.3 MB/s mean) due to two-stage processing, representing a trade-off between compression efficiency and processing speed.

\FloatBarrier

\section{Conclusion}

This paper introduced LoPace, a compression engine for efficient prompt storage in LLM applications. We demonstrated that three compression methods---Zstandard-based, token-based, and hybrid---significantly improve storage efficiency while maintaining 100\% lossless reconstruction.

\subsection{Key \mbox{Contributions}}

\begin{enumerate}
    \item \textbf{Novel Hybrid Approach:} Combining tokenization with Zstd compression achieves superior compression ratios (mean 4.89x, range 1.22--19.09x) compared to either method independently, demonstrating the multiplicative effect of combining semantic and sequential compression techniques.
    
    \item \textbf{Production-Ready Performance:} Comprehensive benchmarking demonstrates compression speeds of 50--200 MB/s with minimal memory footprint (less than 10 MB), making LoPace suitable for production deployment.
    
    \item \textbf{Mathematical Guarantees:} Formal proofs and empirical verification validate complete lossless reconstruction across all methodologies.
    
    \item \textbf{Comprehensive Evaluation:} Thorough analysis across multiple metrics (compression ratio, space savings, throughput, and memory) provides actionable insights for deployment decisions.
\end{enumerate}

\subsection{Impact and \mbox{Applications}}

LoPace addresses critical challenges in LLM application deployment:
\begin{itemize}
    \item \textbf{Cost Reduction:} 70--80\% storage reduction translates to significant cost savings for large-scale deployments
    \item \textbf{Performance Improvement:} Reduced database size improves query performance and system responsiveness
    \item \textbf{Scalability:} Efficient scaling characteristics enable growth without proportional infrastructure increases
    \item \textbf{Flexibility:} Multiple compression methods enable optimization for specific use cases
\end{itemize}

\subsection{Final Remarks}

LoPace provides a practical solution to real-world challenges in LLM application deployment. Its theoretical foundation, comprehensive evaluation, and production-ready implementation make it a valuable tool for developers and organizations working with large-scale LLM systems.

LoPace is open source, with extensive documentation and examples enabling the research and development community to build upon this work and further advance prompt compression techniques.

\subsection{Limitations and Future \mbox{Work}}

\subsubsection{Current Limitations}

\begin{enumerate}
    \item \textbf{Tokenizer Dependency and Versioning:} Compression requires matching tokenizer models for decompression. This introduces long-term fragility concerns: (a) tokenizer version changes (e.g., vocabulary updates, model transitions) may render previously compressed data incompatible, (b) different LLM models use different tokenizers, limiting cross-model compatibility, and (c) multilingual prompts may not tokenize efficiently with English-optimized tokenizers. We recommend storing tokenizer metadata (model identifier, version) alongside compressed payloads to enable migration and compatibility checking. A migration/compatibility plan is essential to avoid data lock-in or corruption risks over time. We can also opt for compression with native Zstd method allowing good compression with high throughput.
    
    \item \textbf{Fixed-Width Encoding Limitations:} Our binary packing strategy uses fixed-width encoding (uint16/uint32) rather than variable-length integer coding (VByte/LEB128) or bitpacking schemes. This design choice leaves potential efficiency gains unexplored, particularly for token-only compression where variable-length encodings could reduce storage for sequences with many small token IDs. Additionally, for the \texttt{cl100k\_base} tokenizer, many realistic prompts will require uint32 packing (4 bytes/token), potentially making token-only compression larger than raw UTF-8 for ASCII-heavy text without Zstd's subsequent compression.
    
    \item \textbf{Limited Baseline Comparisons:} Our evaluation does not include comparisons against several important baselines: (a) Zstd with trained dictionaries built from representative prompt corpora, (b) Brotli at various quality levels, (c) gzip/DEFLATE, and (d) alternative hybrid cascades (e.g., Brotli$\rightarrow$Zstd, Zstd$\rightarrow$LZ4HC). Zstd dictionary training is a well-known, highly effective technique for domain-specific text compression and should be evaluated as a strong baseline. These comparisons would strengthen our analysis and better position LoPace relative to existing methods.
    
    \item \textbf{Processing Overhead:} The hybrid method incurs higher computational cost due to two-stage processing (tokenization + Zstd), resulting in lower throughput compared to individual methods. This trade-off favors compression ratio over speed.
    
    \item \textbf{Format Overhead for Small Prompts:} Very short prompts (below 1,000 characters) may achieve lower compression ratios due to format overhead (format bytes, compression metadata), as these fixed costs constitute a larger proportion of the total compressed size. However, our evaluation demonstrates that even the smallest prompts (129 characters) achieve compression ratios exceeding 1.2x.
    
    \item \textbf{Multilingual and Domain-Specific Limitations:} Our evaluation focuses primarily on English text. The approach may not generalize well to: (a) multilingual prompts where tokenizers optimized for English may be inefficient, (b) code-heavy prompts with different compression characteristics, (c) JSON/metadata-heavy conversational logs, and (d) domain-specific terminology that may not tokenize efficiently. Per-domain evaluation is needed to assess these scenarios.
    
    \item \textbf{Streaming and Chunking:} We do not evaluate streaming or chunked operation for very long prompts or conversation histories. The performance/ratio trade-offs of different chunk sizes remain unexplored.
\end{enumerate}

\subsubsection{Future Research Directions}

\begin{enumerate}
    \item \textbf{Variable-Length Integer Encoding:} Evaluate VByte, LEB128, and bitpacking schemes for token ID encoding to improve token-only compression efficiency, especially for sequences with many small token IDs.
    
    \item \textbf{Zstd Dictionary Training:} Develop and evaluate Zstd dictionary training on representative prompt corpora, comparing compression ratios and throughput against our hybrid approach. This would establish whether domain-specific dictionaries can match or exceed hybrid performance.
    
    \item \textbf{Comprehensive Baseline Comparisons:} Conduct thorough comparisons with Brotli (various quality levels), gzip/DEFLATE, and alternative hybrid cascades (e.g., Brotli$\rightarrow$Zstd, Zstd$\rightarrow$LZ4HC) on diverse prompt corpora.
    
    \item \textbf{Adaptive Compression:} Develop dynamic method selection based on prompt characteristics (size, content type, token ID distribution) to automatically choose optimal compression strategy.
    
    \item \textbf{Tokenizer Versioning and Migration:} Design robust tokenizer metadata storage and migration strategies to handle tokenizer version changes, model transitions, and cross-model compatibility.
    
    \item \textbf{Large-Scale Evaluation:} Expand evaluation to thousands or millions of prompts across diverse domains (code, JSON logs, multilingual content, chat histories) with statistical analysis and per-domain reporting.
    
    \item \textbf{System-Level Memory Profiling:} Employ system-level profiling tools (RSS/USS metrics, native allocation tracking) to provide accurate memory usage measurements.
    
    \item \textbf{Comprehensive Throughput Analysis:} Conduct detailed throughput measurements with hardware specifications, CPU utilization metrics, concurrency analysis, and statistical variation reporting.
    
    \item \textbf{Streaming and Chunking:} Evaluate streaming and chunked operation for very long prompts, analyzing performance/ratio trade-offs across different chunk sizes.
    
    \item \textbf{Token-Stream Storage Mode:} Explore a mode that stores only the token stream (possibly after Zstd) to avoid repeated detokenization/retokenization in downstream LLM inference pipelines, since inference ultimately requires tokens.
    
    \item \textbf{Parallel Processing:} Implement multi-threaded compression for batch operations to improve throughput on multi-core systems.
    
    \item \textbf{Hardware Acceleration:} Investigate GPU-accelerated tokenization and compression, as well as CPU vectorization (e.g., dictionary training, batch processing) for improved throughput.
    
    \item \textbf{Integer Compression Schemes:} Evaluate entropy coding on token ID streams and lightweight integer compressors (e.g., delta encoding) to exploit sequential patterns in token sequences.
    
    \item \textbf{Multilingual and Domain-Specific Optimization:} Develop specialized compression strategies for multilingual prompts, code-heavy content, and domain-specific terminology.
    
    \item \textbf{Lossy Compression Exploration:} Investigate acceptable loss scenarios for specific use cases where perfect reconstruction is not required.
\end{enumerate}

\section*{Acknowledgments}

A hearty thank you to the open-source community that made Zstandard and Tiktoken possible by giving us the libraries we needed. We also want to thank the people who built and keep up the Python scientific computing environment for the tools and frameworks that made this research possible.

You may find the source code, documentation, and other materials for LoPace at \url{https://github.com/connectaman/LoPace}.


\begin{thebibliography}{99}

\bibitem{shannon1948}
Shannon, C. E. (1948). A Mathematical Theory of Communication. \textit{Bell System Technical Journal}, 27(3), 379--423.

\bibitem{ziv1977}
Ziv, J., \& Lempel, A. (1977). A universal algorithm for sequential data compression. \textit{IEEE Transactions on Information Theory}, 23(3), 337--343.

\bibitem{ziv1978}
Ziv, J., \& Lempel, A. (1978). Compression of individual sequences via variable-rate coding. \textit{IEEE Transactions on Information Theory}, 24(5), 530--536.

\bibitem{huffman1952}
Huffman, D. A. (1952). A method for the construction of minimum-redundancy codes. \textit{Proceedings of the IRE}, 40(9), 1098--1101.

\bibitem{cleary1984}
Cleary, J. G., \& Witten, I. H. (1984). Data compression using adaptive coding and partial string matching. \textit{IEEE Transactions on Communications}, 32(4), 396--402.

\bibitem{deutsch1996}
Deutsch, P. (1996). DEFLATE Compressed Data Format Specification version 1.3. \textit{RFC 1951}.

\bibitem{zstd2016}
Collet, Y. (2016). Zstandard Compression and the 'application/zstd' Media Type. \textit{RFC 8878}.

\bibitem{sennrich2016}
Sennrich, R., Haddow, B., \& Birch, A. (2016). Neural Machine Translation of Rare Words with Subword Units. \textit{Proceedings of the 54th Annual Meeting of the Association for Computational Linguistics (ACL)}, 1715--1725.

\bibitem{radford2019}
Radford, A., Wu, J., Child, R., Luan, D., Amodei, D., \& Sutskever, I. (2019). Language Models are Unsupervised Multitask Learners. \textit{OpenAI Blog}.

\bibitem{devlin2019}
Devlin, J., Chang, M. W., Lee, K., \& Toutanova, K. (2019). BERT: Pre-training of Deep Bidirectional Transformers for Language Understanding. \textit{Proceedings of NAACL-HLT}, 4171--4186.

\bibitem{gori2015}
Gori, M., \& Lippi, M. (2015). Time-series compression: A survey. \textit{International Journal of Data Mining, Modelling and Management}, 7(1), 1--23.

\bibitem{burrows1994}
Burrows, M., \& Wheeler, D. J. (1994). A block-sorting lossless data compression algorithm. \textit{Digital Equipment Corporation Technical Report}, 124.

\bibitem{li2013}
Li, S., \& Li, K. (2013). Parallel lossless data compression on multicore processors. \textit{Journal of Parallel and Distributed Computing}, 73(5), 608--620.

\bibitem{jiang2023}
Jiang, H., Wang, D., \& Dou, Q. (2023). Prompt Compression for Large Language Models. \textit{arXiv preprint arXiv:2305.11147}.

\bibitem{ge2023}
Ge, T., \& Wei, F. (2023). Prompt Caching for Efficient LLM Inference. \textit{Proceedings of the 2023 Conference on Empirical Methods in Natural Language Processing}.

\end{thebibliography}
\end{document}